\documentclass[aip,jcp,amsmath,amssymb,longbibliography,floatfix,reprint]{revtex4-2}
\usepackage{mathrsfs}
\usepackage[mathcal]{eucal}
\usepackage{graphicx}
\usepackage{dcolumn}
\usepackage{bm}

\usepackage[utf8]{inputenc}
\usepackage[T1]{fontenc}
\usepackage{mathptmx}
\usepackage{etoolbox} 
\usepackage[caption=false]{subfig}

\PassOptionsToPackage{square,comma,sort&compress,doi=false,isbn=false,issn=false,url=false}{natbib}
\usepackage{anyfontsize}
\usepackage{lipsum}
\usepackage{physics}
\usepackage[dvipsnames]{xcolor}
\usepackage{import}
\usepackage{xifthen}
\usepackage[pdftex,plainpages=false,colorlinks=true,citecolor=blue,linkcolor=blue,urlcolor=blue,filecolor=green,bookmarksopen=true]{hyperref}

\renewcommand{\i}{i}
\newcommand{\Zmath}{\mathbb{Z}}
\newcommand{\oper}[1]{\hat{#1}}
\renewcommand{\H}{\oper{H}}
\newcommand{\Z}{\mathcal{Z}}
\renewcommand{\L}{\oper{{L}}}
\newcommand{\1}{\oper{I}}
\newcommand{\E}[2]{\oper{E}^{#1}_{#2}}
\newcommand{\nn}{\nonumber \\}
\newcommand{\commu}[2]{\comm{#1 \,}{#2}} 
\newcommand{\mmax}{\overline{m}}

\newcommand{\V}[1]{
	\ifthenelse{\isempty{#1}}%
	{\oper{V}}
	{\oper{V}_{#1}}
}
\newcommand{\K}[1]{
	\ifthenelse{\isempty{#1}}%
	{\oper{K}}
	{\oper{K}_{#1}}
}
\newcommand{\m}[2]{%
	\ifthenelse{\isempty{#1}}%
	{\vb*{m}^{\hspace{-0.2mm}\scriptscriptstyle{#2}}}
	{m_{\hspace{-0.1mm}\scriptscriptstyle{#1}}^{\hspace{-0.1mm}\scriptscriptstyle{#2}}}
}
\newcommand{\prob}[2]{%
	\ifthenelse{\isempty{#2}}%
	{\mathrm{P}\hspace{-0.05cm}\left(#1\right)}
	{\mathrm{P}\hspace{-0.05cm}\left(#1\middle|#2\right)}
}

\newcommand{\sub}[1]{_{\hspace{-0.01cm}\scriptscriptstyle{#1}}}

\makeatletter
\def\@email#1#2{%
 \endgroup
 \patchcmd{\titleblock@produce}
  {\frontmatter@RRAPformat}
  {\frontmatter@RRAPformat{\produce@RRAP{*#1\href{mailto:#2}{#2}}}\frontmatter@RRAPformat}
  {}{}
}%
\makeatother
\begin{document}

\preprint{AIP/123-QED}

\title{Path Integral Monte Carlo in the Angular Momentum Basis for a Chain of Planar Rotors}
\author{Estêvão de Oliveira}
\affiliation{Department of Physics \& Astronomy, University of Waterloo, Waterloo, Ontario N2L3G1, Canada}

\author{Muhammad Shaeer Moeed}
\affiliation{Department of Chemistry, University of Waterloo, Waterloo, Ontario N2L3G1, Canada}

\author{Pierre-Nicholas Roy}
\affiliation{Department of Chemistry, University of Waterloo, Waterloo, Ontario N2L3G1, Canada}
\email{pnroy@uwaterloo.ca}

\date{\today}

\begin{abstract}
We introduce a Path Integral Monte Carlo (PIMC) approach that uses the angular momentum  representation for the description of interacting rotor systems. Such a choice of representation allows the calculation of momentum properties without having to break the paths.
The discrete nature of the momentum basis also allows the use of rejection-free Gibbs sampling techniques. 
To illustrate the method, we study the collective behavior of $N$ confined planar rotors with dipole-dipole interactions, a system known to exhibit a quantum phase transition from a disordered to an ordered state at zero temperature. Ground state properties are obtained using the Path Integral Ground State (PIGS) method.
We propose a Bond-Hamiltonian decomposition for the high temperature density matrix factorization of the imaginary time propagator. We show that \textit{cluster-loop} type moves are necessary to overcome ergodicity issues and to achieve efficient Markov Chain updates. Ground state energies and angular momentum properties are computed and compared with Density Matrix Renormalization Group (DMRG) benchmark results. In particular, the derivative of the kinetic energy with respect to the interaction strength estimator is presented as a successful order parameter for the detection of the quantum phase transition. 
\end{abstract}

\maketitle

\section{\label{sec:Introduction}Introduction}

The theoretical description of collections of interacting confined molecules with rotational degrees of freedom is of high relevance to the field of chemical physics.  A number of recent experiments have revealed interesting collective phenomena associated with correlated molecular rotations. For instance, water molecules confined in crystals of beryl\cite{gorshunov2016incipient,kolesnikov2016quantum} and cordierite \cite{belyanchikov2022single} can exhibit ferroelectricity, tunneling, and ordering behavior.
When confined to single-wall carbon nanotubes, a water chain was observed to undergo a thermal quasiphase transition from an ordered to disordered phase.\cite{ma2017quasiphase}
Lattices of water molecules confined in C$_{60}$ have been shown to display an enhanced permittivity at low temperatures,\cite{aoyagi2014cubic} a phenomenon associated with dipole ordering.

The theoretical modeling of confined molecular systems is essential for the prediction of their properties. Due to the quantum mechanical nature of confined molecules, Exact Diagonalization (ED) is the approach of choice and be can applied for assemblies containing a few molecules such as short chains of H$_2$O$@$C$_{60}$ endofullerenes.\cite{felker2017accurate,felker2017electric} 
Basis truncation techniques can be used to treat longer chains such as HF$@$C$_{60}$ model enfullerene peapods.\cite{halverson2018quantifying} However, due to the exponential scaling of the computational cost of ED, the approach is limited to system containing a small number of interacting molecules. Longer chains containing hundreds of confined molecules can be studied in the ground state using the Density Matrix Renormalization Group (DMRG) method,\cite{white1992density} 
using a Matrix Product State (MPS)\cite{schollwock2011density} ansatz for the many-body ground state wavefunction. This approach has been used to study 1-d chains of dipolar linear molecules,\cite{iouchtchenko2018ground} rotating water molecules\cite{serwatka2022ground,Serwatka2023}, water chains in carbon nanotubes,\cite{serwatka2022ferroelectric}  various endofullerence peapods,\cite{serwatka2023endo} and chains of dipolar planar rotors.\cite{Serwatka2024}
The MPS form of the ground state wavefunction is however limited to one-dimensional chains and alternate approaches are therefore required. Some initial efforts have shown that the Multi Configuration Time Dependent Hartree (MCTDH) with its multi-layer tree network structure of the wavefunction is a potential avenue.\cite{mainali2021comparison} Recurrent Neural network have also been shown as another promising model for the ground state wavefunction of dipolar planar rotors.\cite{serwatka2024ground}
Another approach to simulate many-body quantum systems is based on Quantum Monte Carlo (QMC) techniques.\cite{ceperley1986quantum,gubernatis2016quantum} In particular, the Path Integral Monte Carlo (PIMC) method\cite{ceperley1995path} is a powerful and general tool amenable to one-, two-, and three-dimensional systems. Although originally formulated in Cartesian coordinates, PIMC can readily by adapted to curved spaces and rotor systems.\cite{marx1999path}
For ground state properties, the Path Integral Ground State (PIGS) formulation\cite{sarsa2000path,yan2017path} is an approach of choice. 
Systems of dipolar rotors as well as water chains have been successfully simulated using PIGS.\cite{abolins2011ground,abolins2013erratum,abolins2018quantum,sahoo2020path,sahoo2021path} 

Recent work has shown that the PIMC/PIGS approach for rotors can be formulated in discrete angular representation.\cite{Zhang2024,Moed2024} 
The resulting imaginary time paths on grids lead to path sums rather path integrals. In turn, each configuration of the path therefore has a finite probability in contrast to continuous position path integrals. 
This allows for the implementation of powerful Gibbs sampling\cite{casella1992explaining,gelfand2000gibbs} schemes for the construction of Markov chains. 
Here, we propose a formulation of rotor PIMC in the discrete angular momentum representation.
Our approach allows the {\em direct} calculation of angular momentum properties without having to break imaginary time paths as usually done when computing momentum properties that are non-local in the position representation.\cite{perez2011improving,cendagorta2018open} 
To test our approach, we simulate a chain of dipolar planar rotors, a system that has been shown to undergo a $(1+1)d$  $Z_2$ symmetry breaking Quantum Phase Transition (QPT) from an ordered to a disordered phase.\cite{Serwatka2024} 
 
The remainder of this paper is organized as follows: in Sec. \ref{sec:Theory} we introduce the dipolar rotor system and describe our the PIGS formalism. We report numerical simulation results in Sec. \ref{sec:Results} and discuss our findings. We provide concluding remarks in Sec. \ref{sec:Conclusions} along with a perspective for future work.

\section{Theory}
\label{sec:Theory}

\subsection{Planar Rotors in the Angular Momentum Basis Representation}
\label{sec:planar_rotors}
\begin{figure}
  \centering
  \includegraphics[width = \columnwidth]{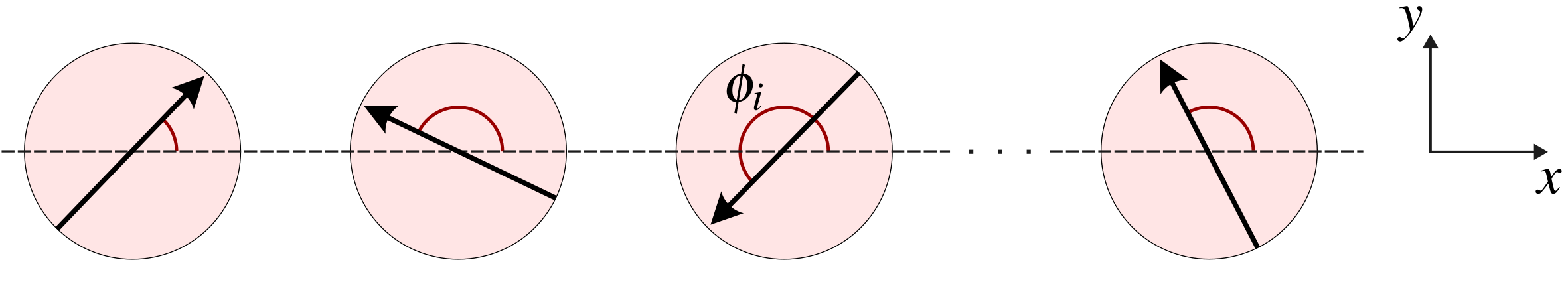}
  \caption{Illustration of the $N$ planar rotor chain in a co-planar arrangement.\cite{Serwatka2024}}
  \label{fig:planar_rotors_chain}
\end{figure}

The system of interest is constituted of $N$ identical planar rotors equally spaced with some distance $r$ in a linear chain, each with angular orientation $\phi$ in the $xy$-plane, and dipole-dipole nearest neighbors interaction (n.n.i.) and open boundary conditions (OBC). The model is depicted in Fig. \ref{fig:planar_rotors_chain}. 
The dimensionless classical Hamiltonian  can be written as
\begin{align}
\label{eq:classical Hamiltonian}
	H = \sum_{i=1}^{N} p_{\phi_i}^2 + g \sum_{\langle ij \rangle} \sin{\phi_i} \sin{\phi_j} -2\cos{\phi_i} \cos{\phi_j},
\end{align} 
where $(\phi_i,p_{\phi_i})$ is the canonically conjugated pair of angle and orbital angular momentum variables, $g$ is the interaction strength normalized by the moment of inertia, and is proportional to $1/r^3$.

A thorough quantization process of this system involves the replacement of the variable $\phi_i$ by the smooth periodic functions $\sin{\phi_i}$ and $\cos{\phi_i}$. Then, the phase space can be quantized in terms the self-adjoint generators $\left\{\L_{i},\E{\pm}{i}\right\}$, that form the algebra\cite{KLEINERT199,KOWALSKI2002109,Kastrup_2006,orr2024operator} 
\begin{align}
\label{eq:algebra_L,Ep,Em_ops}
	\commu{\L_i}{\E{\pm}{i}} = \pm \E{\pm}{i}, \qquad
	\commu{\E{\pm}{i}}{\E{\mp}{i}} = 0. 
\end{align}
The action of angular momentum operator $\L_i$ and the raising and lowering operators $\E{\pm}{i}$ in the angular momentum basis $\{\ket{m_i}, \forall m_i \in \Zmath\}$ are given by\cite{eh_ek_2008}
\begin{align}
  \L_i \ket{m_i} =  m_i \ket{m_i},
  \qquad
  \E{\pm}{i} \ket{m_i} = \ket{m_i \pm 1} \label{eq:action_L,Ep,Em_ops}.
\end{align}
The quantum Hamiltonian operator can now be written as
\begin{align}
    \label{eq:hamiltonian_op_quantum_system}
    \H = \sum_{i=1}^N \K{i} + g \sum_{\expval{ij}} \V{ij},
\end{align}
where $\K{i} \equiv \L_i^2$ is the kinetic energy operator of the $i$-th quantum rotor, and
\begin{align}
	\label{eq:potential_energy_op}
	\V{ij} = - \frac{1}{4} \left( 3\E{+}{i}\E{+}{j}+\E{+}{i}\E{-}{j}+\E{-}{i}\E{+}{j}+3\E{-}{i}\E{-}{j} \right),
\end{align}
is the potential energy operator representing the interaction between the $i$-th and $j$-th sites.

The angular momentum basis forms an infinite discrete orthonormal set, so in order to achieve a practical computational implementation of the system it is necessary to truncate the space generated by $\{\ket{m_i}\}$. Setting a maximum value of angular momentum, defined as $\mmax$, the $(2\mmax+1)$-dimensional basis set is $\{\ket{m_i}: m_i \in [-\mmax,+\mmax]\}$. 
It is worth mentioning that, although this could be achieved in many ways, all of those would introduce errors or inconsistencies since the algebraic relations of Eq. \eqref{eq:algebra_L,Ep,Em_ops} are well defined only for an infinite dimensional space. 
This kind of problem has been reported for bosonic systems in a finite basis\cite{Somma_2003,castillo2023,Hanada_2023}. 
Here, the most straightforward way is to simply set 
\begin{align}
	\E{\pm}{} \ket{\pm\mmax} = 0. \label{eq:E_pm_action_mmax_lambda_0}
\end{align}

\subsection{Path Integral Ground State for planar rotors}
\label{sec:PIGS_planar_rotors}

In the PIGS formulation one starts with some trial state vector $\ket{\Psi_T}$, 
non-orthogonal to the true ground state.
That trial state is propagated in imaginary time $\beta$ according to,
\begin{align}
  \ket{\Psi_{\beta}} =& e^{-\frac{\beta}{2} \oper{H}} \ket{\Psi_T} .
\end{align}
The ground state is obtained in the $\beta \rightarrow \infty$ limit and the expectation value of a physical observable $\oper{O}$ is estimated from,
\begin{align}
  \label{eq:expectation_PIGS}
  \expval{\oper{O}}_{\textbf{gs}}
  = \lim_{\beta \rightarrow \infty}\frac{\mel**{\Psi_{\beta}}{\oper{O}}{\Psi_{\beta}}}{\Z(\beta)} 
  ,
\end{align}
where $\Z(\beta) \equiv \braket{\Psi_{\beta}}$ is a normalization known as the (pseudo)-partition function, in  analogy with the partition function defined in the finite temperature canonical ensemble. It is important to note that
\begin{align}
  \label{eq:pseudo_partition_function}
  \Z(\beta)  =
  \bra{\Psi_T} e^{-\frac{\beta}{2} \oper{H}}
  e^{-\frac{\beta}{2} \oper{H}} \ket{\Psi_T}
  = \tr[e^{-\beta \oper{H}} \ketbra{\Psi_T}],
\end{align}
where the explicit calculation of the $(2\mmax+1)^N$-dimensional density matrix $\oper{\rho}(\beta) \equiv e^{-\beta \oper{H}}$ is intractable for large systems due to the exponential scaling of the Hilbert space.

We first use the semigroup (factorization) property of the thermal density operator to obtain,
\begin{align}
  \label{eq:partition_function_PIMC}
  \Z(\beta)
  =& \sum_{\m{}{1}} \mel{\m{}{1}}{ \left[ e^{-\frac{\beta}{L} \H} \right]^L}{\m{}{1}} \nn
	=& \sum_{\{\m{}{l}\}_{L}} \prod_{l=1}^{L} \mel{\m{}{l}}{\oper{\rho}(\tau)}{\m{}{l+1}},
\end{align}
where $\tau = \frac{\beta}{L}$ with $L$ even, and $\ket{\m{}{L+1}} = \delta_{\,\Psi_T,\m{}{1}}\ket{\m{}{1}}$ for the PIGS formulation. Resolutions of the identity $\1 = \sum_{\m{}{\alpha}} \ketbra{\m{}{\alpha}}$ (also know as ``\textit{beads}'' in the context of PIGS) were inserted in between all the $L$ factors of $e^{-\tau \H}$. In the notation adopted through this paper,
\begin{align}
  \sum_{\{\m{}{l}\}_{L}} \equiv \sum_{\m{}{1}} \sum_{\m{}{2}} \sum_{\m{}{3}} \dots \sum_{\m{}{L}},
\end{align}
where the superscripts are used to differentiate between different multi-particle states $\ket{\m{}{}} = \ket{\m{1}{},\m{2}{},\dots, \m{N}{}}$, with the subscript designating the rotors.
\begin{equation}
    \label{eq:trial_state}
    \ket{\Psi_T} \equiv \ket{\vb{0}} = \bigotimes_{i=1}^{N} \ket{0}.
\end{equation}
This corresponds to a trial function where all the rotors in the $m=0$ angular momentum state . 

To use \eqref{eq:partition_function_PIMC} in PIGS calculations, the thermal density operator must be approximated via Trotter-like factorization as discussed below.

\subsection{\label{sec:bond_H_decomp}Bond-Hamiltonian Decomposition}
\begin{figure}
  \centering
  \includegraphics[width =\columnwidth]{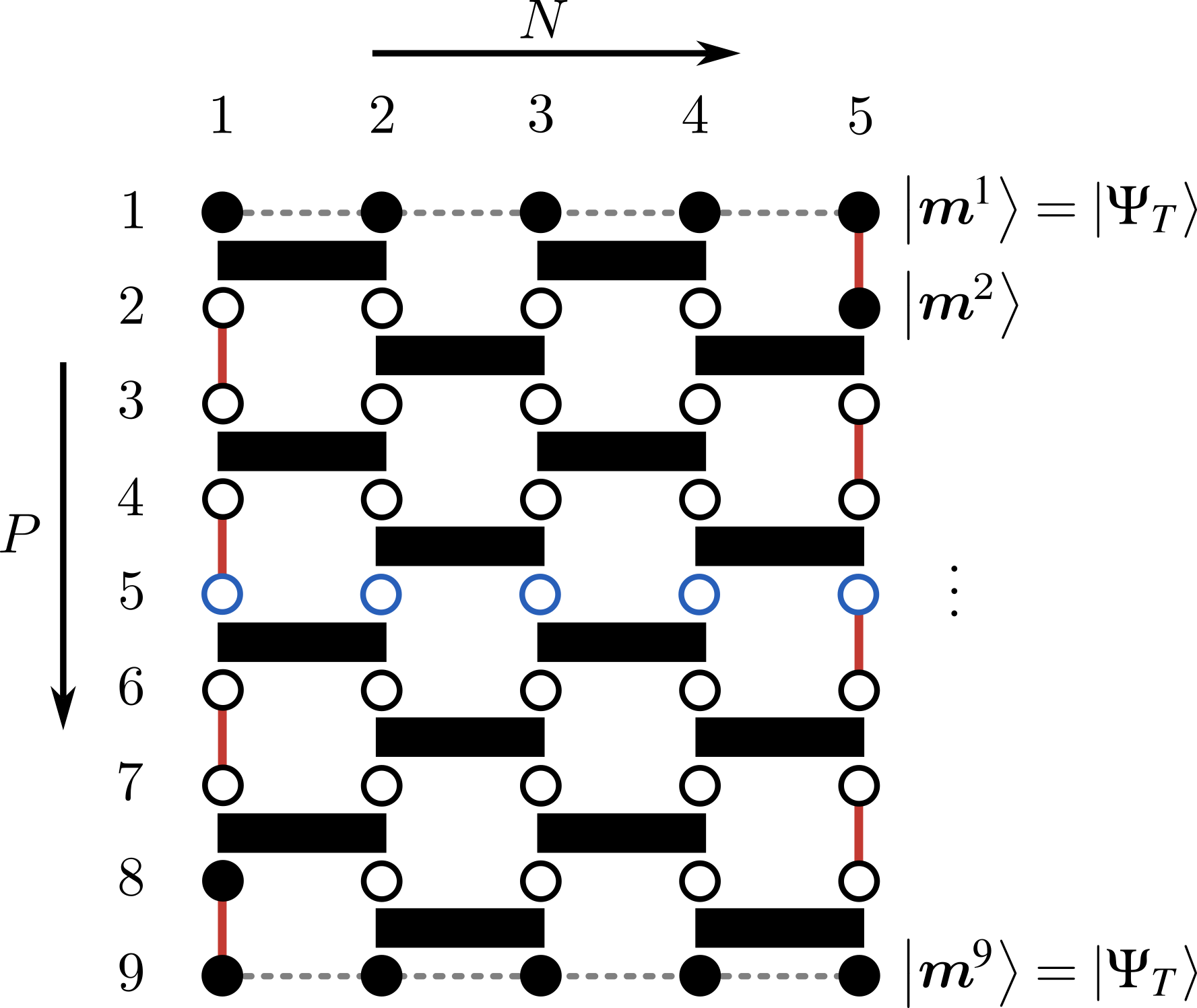}
  \caption{Graphical representation of the ``checker-board'' grid defined in Eq. \eqref{eq:part_function_H_bond_odd_even} for a system of $N=5$ planar rotors and $L=4$ ($P\equiv 2L+1=9$ total beads). The black circles represent the particles (solid for the $m=0$ state and unfilled for a generic state $m$) and the black solid rectangles represent the 2-body density matrix $\rho^{\text{2-B}}_{\text{bond}}(\tau)$. The blue circles indicate the middle bead for $P=L+1=5$. The red solid lines stand for the kinetic energy density matrix operators $e^{- \frac{\tau}{2} \oper{K}_{1}}$ and $e^{- \frac{\tau}{2} \oper{K}_{5}}$ acting on the $1$st and $5$th particle due to the OBC.}
  \label{fig:PIMC_grid_N=5_P=9_bond}
\end{figure}

The total Hamiltonian can be decomposed as
\begin{align}
  \label{eq:total_H_bond_decomp}
  \H 
  = \frac{\K{1}}{2} + \frac{\K{N}}{2} + \sum_{\langle ij\rangle} \H_{ij}  ,
\end{align}
where
\begin{align}
  \label{eq:bond-H}
  \H_{ij}  \equiv \frac{\K{i} + \K{j}}{2} + \V{ij} 
\end{align}
stands for the Hamiltonian of the bond between the  $i$-th and $j$-th particles. Due to the non-commutation between terms $\H_{ij}$ and $\H_{jl}$, it is convenient to group 2-body bond Hamiltonian terms that do not share the same Hilbert space, by defining
\begin{align}
  \label{eq:bond-H_odd_even_def}
    &\H_{\text{odd}} \equiv \H_{12} + \H_{34} + \dots =  \sum_{\substack{i=1\\(i\text{ odd})}}^{N-1} \H_{i,i+1} ,\\
    &\H_{\text{even}} \equiv \H_{23} + \H_{45} + \dots =\sum_{\substack{i=2\\(i\text{ even})}}^{N-1} \H_{i,i+1}  .
\end{align} 
Then, in the Bond-Hamiltonian (BH) decomposition, the density  $\oper{\rho}(\tau) = e^{-\tau \H}$ can be approximated by applying the Trotter formula
\begin{align}
    \label{eq:dens_matrix_tau_bond_H}
    \oper{\rho}(\tau)
    &=e^{-\tau \left( \frac{\K{1}}{2} + \H_{\text{odd}} + \H_{\text{even}} + \frac{\K{N}}{2}\right)}
    \nn
    &\approx 
    \begin{cases}
        e^{-\tau \H_{\text{odd}}}  e^{- \frac{\tau}{2} \K{N}} ~\cdot~  e^{- \frac{\tau}{2} \K{1}}e^{-\tau \H_{\text{even}}}, & \text{ if $N$ odd}, \\
        e^{-\tau \H_{\text{odd}}} ~\cdot~ e^{- \frac{\tau}{2} \K{1}} e^{-\tau \H_{\text{even}}}  e^{- \frac{\tau}{2} \K{N}}, & \text{ if $N$ even},
    \end{cases}
\end{align}
which corresponds to the usual Symmetric Primitive (SP) decomposition\cite{ceperley1995path}
\begin{align}
    \oper{\rho}(\tau)
    \approx  e^{-\frac{\tau }{2} \K{}} e^{-\tau \V{}} e^{-\frac{\tau }{2} \K{}} ,
\end{align}
when another Trotter expansion is applied to the $e^{-\tau \H_{\text{odd}}}$ and $e^{-\tau \H_{\text{even}}}$ density operators in Eq. \eqref{eq:dens_matrix_tau_bond_H}. 
We adopt a notation where the terms $e^{- \frac{\tau}{2} \K{1}}$ and $e^{- \frac{\tau}{2} \K{N}}$ due to the OBC are absorbed in the other density operators.
Matrix elements of the operator defined in Eq. \eqref{eq:dens_matrix_tau_bond_H} can be written as,
\begin{align}
    \label{eq:density_matrix_N_body_bead}
    &\mel{\m{}{l}}{\oper{\rho}(\tau)}{\m{}{l+1}} \approx
    \nn 
    &=\mel{\m{}{l}}{e^{-\tau \H_{\text{odd}}} \cdot e^{-\tau \H_{\text{even}}}}{\m{}{l+1}}
    \nn 
    &= \sum_{\m{}{\alpha\sub{l}}} \mel{\m{}{l}}{ e^{-\tau \H_{\text{odd}}}}{\m{}{\alpha\sub{l}}} 
    \mel{\m{}{\alpha\sub{l}}}{e^{-\tau \H_{\text{even}}}}{\m{}{l+1}},
\end{align}
Now, inserting Eq. \eqref{eq:density_matrix_N_body_bead} into Eq. \eqref{eq:partition_function_PIMC}, reorganizing the sums and relabeling the terms,
we have 
\begin{align}		
\label{eq:partition_function_PIMC_trotter_Heven_Hodd_ini}
    \Z(\beta) 
    \approx \sum_{\{\m{}{p}\}_{2L}} \Pi\left(\m{}{1},\m{}{2L+1};\tau\right),
\end{align}
where we define
\begin{align}
    \label{eq:propagator}
    \Pi\left(\m{}{1},\m{}{2L+1};\tau\right) \equiv \prod_{p=1}^{2L} \mel*{\m{}{p}}{\oper{\rho}_p(\tau)}{\m{}{p+1}},
\end{align}
and
\begin{equation}
    \label{eq:def_rho_tau_op}
    \oper{\rho}_p(\tau) =
    \left\{
    \begin{aligned}
        e^{ -\tau \H_{\text{odd}}} & \text{, for $p$ odd,} \\
        e^{ -\tau \H_{\text{even}}} & \text{, for $p$ even.} 			
    \end{aligned}
    \right.
\end{equation}
Again, it is implicit that $\ket{\m{}{2L+1}} = \delta_{\Psi_T,\m{}{1}}\ket{\m{}{1}}$.
The matrix elements in Eq. \eqref{eq:propagator} can each still be decomposed into the product of the matrix elements of 2-body density matrices
\begin{align}
    \mel**{\m{}{p}}{\oper{\rho}_p(\tau)}{\m{}{p+1}}
    =& \mel{\m{}{p}}{\prod_{i \in \mathcal{A}_p} e^{ -\tau \H_{i,i+1}} }{\m{}{p+1}}\nn
    =& \prod_{i \in \mathcal{A}_p} \mel{\m{i}{p},\m{i+1}{p}}{}{\m{i}{p+1},\m{i+1}{p+1}} ,
\end{align}
for the set $\mathcal{A}_p$ off all odd (even) numbers in the interval $\{1,N-1\}$ if $p$ is an odd (even) number. 
The double bar notation $||$ represents the matrix element being taken with respect to the the $(2\mmax+1)^2$-dimensional 2-body density matrix $\oper{\rho}^{\text{2-B}}_{\text{bond}}(\tau) \equiv e^{-\tau \oper{h}}$, which can easily be calculated and stored. In this context, $\oper{h}$ is the $(2\mmax+1)^2$-dimensional Hamiltonian of the bond between two rotors. Now, we can finally approximate the pseudo-partition function of Eq. \eqref{eq:pseudo_partition_function} as
\begin{align}
    \label{eq:part_function_H_bond_odd_even}
    \Z(\beta)
    = \sum_{\{\m{}{p}\}_{2L}}
    \prod_{p=1}^{2L} \prod_{i \in \mathcal{A}_p} \mel{\m{i}{p},\m{i+1}{p}}{}{\m{i}{p+1},\m{i+1}{p+1}},
\end{align}
up to a $\tau$ error, where now the sums can be estimated stochastically through a Markov Chain Monte Carlo (MCMC) algorithm. The definition above leads to the ``checker-board'' grid graphical representation, depicted in Fig. \ref{fig:PIMC_grid_N=5_P=9_bond}. 

\subsection{\label{sec:PIGS_Estimators}PIGS Estimators}
Expectation values can be approximately obtained from their respective estimators. We first define a grid state as
\begin{align}
    \va{M} 
    &= \left(\m{}{1}, \m{}{2}, \dots, \m{}{2L+1}\right)
    \nn
    &= \left(\m{1}{1}, \dots, \m{N}{1},\dots, \m{p}{n}, \dots, \m{1}{2L+1}, \dots, \m{N}{2L+1}\right)
    \nn
    & \equiv \left(M_1,M_2,\dots,M_{N(2L+1)}\right).
\end{align}
In the cases where $\Z(\beta)$ is the quantity being sampled in the MCMC process, the derivation of the estimator involves rewriting Eq. \eqref{eq:expectation_PIGS} in the form\cite{book_manybody, sandvik_review}
\begin{align}
    \label{eq:expectation_PIGS_prob_form}
    \expval{\oper{O}} = \sum_{\{\va{M}\}} \frac{W(\va{M})}{\Z(\beta)} \mathcal{O}(\va{M}),
\end{align}
where the sum is taken over all possible $(2\mmax+1)^{(2L-1)N}$ grid configurations\footnote{Only $2L-1$ beads change their configuration since the first and last are bounded by the trial state $\ket{\Psi_T}$.}, $W(\va{M})$ corresponds to the weight of a certain configuration, given by Eq. \eqref{eq:propagator}, and $\mathcal{O}(\va{M})$ is its contribution to the observable $\oper{O}$.\footnote{The calligraphic notation is used to distinct between operators and their respective estimators.} Hence, ${W(\va{M}})/{\Z(\beta)}$ can be interpreted as the the probability distribution for the set of all possible configurations $\{\va{M}\}$.

\begin{equation}
  \label{eq:bond-O_odd_even}
  \oper{O}_{\text{bonds}} = \oper{O}_{\text{odd}} + \oper{O}_{\text{even}} ,
\end{equation}
for
\begin{equation}
  \label{eq:bond-O_odd_even_def}
  \left\{
  \begin{aligned}
    &\oper{O}_{\text{odd}} \equiv \oper{O}_{12} + \oper{O}_{34} + \dots =  \sum_{i \text{ odd}} \oper{O}_{ij} ,\\
    &\oper{O}_{\text{even}} \equiv \oper{O}_{23} + \oper{O}_{45} + \dots =\sum_{i \text{ even}} \oper{O}_{ij}  .
  \end{aligned}
  \right.
\end{equation}

In the angular momentum basis, all the observables are represented by matrices, which do not commute with the propagator $\oper{\rho}(\tau)$ in most cases. Consequently, observables that cannot meet the n.n.i. BH decomposition representation and be expressed in the form of Eq. \eqref{eq:bond-O_odd_even} introduce additional computational complexity in their calculations. This is because extra beads need to be added to account for operator that involve long range off-diagonal correlations. Thus, those type of operators will not be analyzed here. Now, proceeding with the calculation of the numerator of Eq. \eqref{eq:expectation_PIGS} we have
\begin{align}
    \expval**{\oper{O}}{\Psi_{\beta}}=\expval**{\oper{\rho}({\scriptstyle\frac{\beta}{2}}) \left(\oper{O}_{\text{odd}} + \oper{O}_{\text{even}}\right) \oper{\rho}({\scriptstyle\frac{\beta}{2}})}{\Psi_T}
    ,
\end{align}
and we can calculate the odd and even terms, respectively
\begin{widetext}
\begin{align}
\label{eq:expval_numerator_odd_term}
&\expval**{\oper{\rho}({\scriptstyle\frac{\beta}{2}}) \cdot \oper{O}_{\text{odd}}\cdot \oper{\rho}({\scriptstyle\frac{\beta}{2}})}{\Psi_T} = \nn
&= \expval**{
    \left[\oper{\rho}(\tau)\right]^{\frac{L}{2}}
    \cdot \oper{O}_{\text{odd}} \cdot \oper{\rho}(\tau)   \cdot
    \left[\oper{\rho}(\tau)\right]^{\left(\frac{L}{2}-1\right)}
    }{\Psi_T} 
\nn
&\approx \sum_{\{\m{}{p}\}_{2L}}
    \Pi\left(\m{}{1},\m{}{L+1};\tau\right)
    \mel{\m{}{L+1}}{\oper{O}_{\text{odd}}\cdot e^{-\tau \H_{\text{odd}}}}{\m{}{L+2}} 
    \mel{\m{}{L+2}}{ e^{-\tau \H_{\text{even}}}}{\m{}{L+3}}
    \Pi\left(\m{}{L+3},\m{}{2L+1};\tau\right)
\nn
&= \sum_{\{\m{}{p}\}_{2L}}
    \Pi\left(\m{}{1},\m{}{L+1};\tau\right)
    \mel{\m{}{L+1}}{e^{-\tau \H_{\text{odd}}}}{\m{}{L+2}}
    \mel{\m{}{L+2}}{ e^{-\tau \H_{\text{even}}}}{\m{}{L+3}}
    \Pi\left(\m{}{L+3},\m{}{2L+1};\tau\right)
    \frac{\mel{\m{}{L+1}}{\oper{O}_{\text{odd}}\cdot e^{-\tau \H_{\text{odd}}}}{\m{}{L+2}} }{\mel{\m{}{L+1}}{e^{-\tau \H_{\text{odd}}}}{\m{}{L+2}} }
\nn
&= \sum_{\{\m{}{p}\}_{2L}}
    \Pi\left(\m{}{1},\m{}{2L+1};\tau\right)
    \times
    \mathcal{O}_{\text{odd}}(\tau)
    ,
\end{align}
\end{widetext}
where we define
\begin{align}
    \label{eq:expval_Oodd_def}
    \mathcal{O}_{\text{odd}}(\m{}{};\tau)
    = \frac{
        \mel{\m{}{L+1}}{\oper{O}_{\text{odd}}\cdot e^{-\tau \H_{\text{odd}}}}{\m{}{L+2}}
        }{
        \mel{\m{}{L+1}}{e^{-\tau \H_{\text{odd}}}}{\m{}{L+2}}
        }.
\end{align}
A reweighting process given by the multiplication and division by the term $\mel{\m{}{L+1}}{e^{-\tau \H_{\text{odd}}}}{\m{}{L+2}}$ is performed in the last steps of Eq. \eqref{eq:expval_numerator_odd_term}. This ensures that the weight contribution of an observable $\oper{O}$ is taken into account --- especially for off-diagonal operators --- when the grid configurations contributing to $\Z(\beta)$ are necessarily what is being importance sampled\cite{sandvik_review, book_manybody, Crosson_2021}. This step is crucial to avoid the extra effort of sampling an exponentially enlarged space imposed by \textit{broken world lines}. We can proceed in an analogous manner for the even terms, leading to
\begin{align}
\label{eq:expval_numerator_even_term}
&\expval**{\oper{\rho}({\scriptstyle\frac{\beta}{2}}) \cdot \oper{O}_{\text{even}}\cdot \oper{\rho}({\scriptstyle\frac{\beta}{2}})}{\Psi_T} = \nn
&\hspace{2cm}= \sum_{\{\m{}{p}\}_{2L}}
    \Pi\left(\m{}{1},\m{}{2L+1};\tau\right)
    \times
    \mathcal{O}_{\text{even}}(\tau),
\end{align}
for
\begin{align}
    \label{eq:expval_Oeven_def}
    \mathcal{O}_{\text{even}}(\m{}{};\tau)
    = \frac{
        \mel{\m{}{L}}{ e^{-\tau \H_{\text{even}}} \cdot \oper{O}_{\text{even}}}{\m{}{L+1}} 
        }{
        \mel{\m{}{L+1}}{e^{-\tau \H_{\text{even}}}}{\m{}{L+2}}
        }.
\end{align}
Noticing that
\begin{align}
     \oper{O}_{\text{odd}}\cdot e^{-\tau \H_{\text{odd}}} = \sum_{i \text{ odd}} \oper{O}_{i,i+1}\cdot e^{-\tau \H_{i,i+1}} \prod_{j\neq i} e^{-\tau \H_{j,j+1}} 
\end{align}
terms in the numerator and denominator of Eq. \eqref{eq:expval_Oodd_def} and \eqref{eq:expval_Oeven_def} will cancel out. This leads to
\begin{align}
    \label{eq:expval_Oodd_def_2-body}
    \mathcal{O}_{\text{odd}}
    =
    \sum_{i \text{ odd}}
    \frac{
        \mel{\m{i}{L+1},\m{i+1}{L+1}}{\oper{o} \cdot e^{-\tau \oper{h}}}{\m{i}{L+2},\m{i+1}{L+2}}
        }{
        \mel**{\m{i}{L+1},\m{i+1}{L+1}}{ }{\m{i}{L+2},\m{i+1}{L+2}}
        },
\end{align}
and
\begin{align}
    \label{eq:expval_Oeven_def_2-body}
    \mathcal{O}_{\text{even}}
    =
    \sum_{i \text{ even}}
    \frac{
        \mel{\m{i}{L},\m{i+1}{L}}{e^{-\tau \oper{h}} \cdot \oper{o} }{\m{i}{L+1},\m{i+1}{L+1}}
        }{
        \mel**{\m{i}{L},\m{i+1}{L}}{ }{\m{i}{L+1},\m{i+1}{L+1}}
        }
        ,
\end{align}
for $\oper{o}$ the $(2\mmax+1)^2$-dimensional operator version of $\oper{O}$ for the bond between any two rotors. 
Finally, putting Eqs. \eqref{eq:expval_Oodd_def} and \eqref{eq:expval_Oeven_def} together back in Eq. \eqref{eq:expectation_PIGS} and defining $\mathcal{O} \equiv \mathcal{O}_{\text{odd}} + \mathcal{O}_{\text{even}}$, the expectation value in the regime $\beta \rightarrow \infty$ for the physical observable $\oper{O}$ in the form of Eq. \eqref{eq:expectation_PIGS_prob_form} becomes
\begin{align}
  \label{eq:expectation_PIGS_trotterized}
  \expval{\oper{O}}_{\textbf{gs}}
  \approx \sum_{\{\m{}{p}\}_{2L}} \frac{ \Pi\left(\m{}{1},\m{}{2L+1};\tau\right)}{\Z(\beta)} \times \mathcal{O}(\m{}{};\tau)
  ,
\end{align}
up to an error in $\tau$. 
For the case where $\oper{O}$ is a diagonal operator, Eqs. \eqref{eq:expval_Oodd_def_2-body} and \eqref{eq:expval_Oeven_def_2-body} simply reduce to
\begin{align}
  \label{eq:O_est_diagonal}
  \mathcal{O} = \sum_{i=1}^N \mel**{\m{i}{L+1}}{\oper{O}}{\m{i}{L+1}}
  ,
\end{align}
that is, the expectation value of $\oper{O}$ taken with respect to the middle bead. The sum in Eq. \eqref{eq:expectation_PIGS_trotterized} over all the possible $(2\mmax+1)^{2LN}$ grid states will first be carried out exactly using Numerical Matrix Multiplication (NMM)\cite{nmm_ref} to compute the ground state energy of a small system ($N=3$ rotors), and compared with the Exact Diagonalization (ED) results. This will assist the analysis of the $\tau$ convergence of energy estimators (see Sec. \ref{sec:tau_convergency}). Subsequently, when analyzing the collective behavior of larger systems ($N=150$ rotors), the sum in Eq. \eqref{eq:expectation_PIGS_trotterized} will be carried on stochastically (Sec. \ref{sec:Results}). The estimators of interest will be the kinetic energy
\begin{equation}
    \label{eq:kinetic_estimator}
    \mathcal{K} \equiv \sum_{i=1}^N \left(\m{i}{L+1}\right)^2 \,,
\end{equation}
the total angular momentum
\begin{equation}
    \label{eq:L2_estimator}
    \mathcal{L} \equiv \sum_{i=1}^N \m{i}{L+1} \,,
\end{equation}
the total angular momentum squared $\mathcal{L}^2$,
and the potential energy
\begin{align}
    \label{eq:potential_estimator}
    \mathcal{V} \equiv &
    \sum_{i \text{ odd}}
    \frac{
        \mel{\m{i}{L+1},\m{i+1}{L+1}}{\oper{v} \cdot e^{-\tau \oper{h}}}{\m{i}{L+2},\m{i+1}{L+2}}
        }{
        \mel**{\m{i}{L+1},\m{i+1}{L+1}}{ }{\m{i}{L+2},\m{i+1}{L+2}}
        }
        \nn
    & +
    \sum_{i \text{ even}}
    \frac{
        \mel{\m{i}{L},\m{i+1}{L}}{e^{-\tau \oper{h}} \cdot \oper{v} }{\m{i}{L+1},\m{i+1}{L+1}}
        }{
        \mel**{\m{i}{L},\m{i+1}{L}}{ }{\m{i}{L+1},\m{i+1}{L+1}}
        }
    \,,
\end{align}
for $\oper{v}$ the $(2\mmax+1)^2$-dimensional operator of Eq. \eqref{eq:potential_energy_op}. 
Also, we will analyze the derivative of the Kinetic energy with respect to $g$
\begin{equation}
    \label{eq:dKdg_estimator}
      \dv{\expval{\K{}}}{g} = \beta \left[\expval{\mathcal{K} \cdot \mathcal{V}_{\text{all}}} - \expval{\K{}}\cdot\expval{\V{}}_{\text{all}} \right]
\end{equation}
where in this case $\mathcal{V}_{\text{all}}$ stands for  operator $\oper{V}$ acting on all the beads and $\expval{\V{}}_{\text{all}} \equiv \expval{\mathcal{V}_{\text{all}}}$ (see the Supplementary Material).

\subsection{\label{sec:gibbs_samp}Gibbs Sampling}
We now turn our attention to the importance sampling process required to  stochastically compute the physical properties given in Eq. \eqref{eq:expectation_PIGS_trotterized}.
The discrete nature of the angular momentum representation makes the advantageous rejection-free Gibbs sampling (also known as Heat-Bath sampling) a natural choice.\cite{casella1992,book_manybody} We will show that a certain type of collective updates --- which we will call \textit{cluster-loops} --- although still local, are necessary to overcome ergodicity issues and to achieve efficient sampling.

\begin{figure}[htb]
    \centering
    \subfloat[]{\includegraphics[width = 0.3\columnwidth]{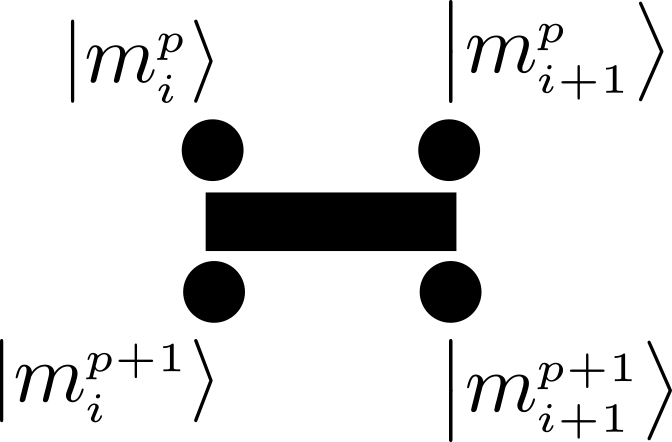}
    \label{fig:building_block}}
    \qquad
    \subfloat[]{\includegraphics[width = 0.3\columnwidth]{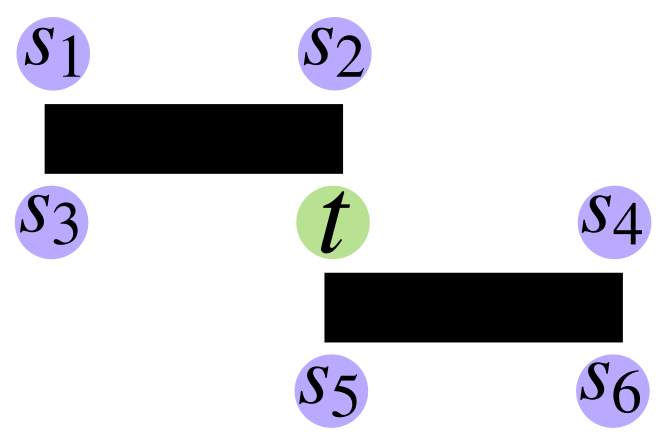}\label{fig:two_cluster_interaction}}
    \caption{Diagrammatic representation of: (a) The matrix element (cluster) of the 2-body density matrix $\oper{\rho}^{\text{2-B}}_{\text{bond}}(\tau)$ for some coordinate particle-bead $(i,p)$ of the grid; (b) The interaction of two clusters where the target particle $\ket{m_t}$ (light green) interacts with the six remaining ones $\{\ket{m_k} : k \in [1,6]\}$ (light purple).}
\end{figure}

It will prove useful to adopt a diagrammatic notation. We show in Fig. \ref{fig:building_block} the building block of the sampling process, referred to as a \textit{cluster}, that consists of the matrix elements $ \mel**{\m{i}{p},\m{i+1}{p}}{ }{\m{i}{p+1},\m{i+1}{p+1}}$ of the 2-body density matrix $\oper{\rho}^{\text{2-B}}_{\text{bond}}(\tau)$.
We can also see from Fig. \ref{fig:PIMC_grid_N=5_P=9_bond} that each particle (black circle) belongs to two different clusters, and hence directly interacts with six other particles in the PIGS grid, as depicted in Fig. \ref{fig:two_cluster_interaction}. During the sampling process, a single MC step represents the full update of an entire grid state $\va{M}_{\text{old}} \longrightarrow \va{M}_{\text{new}}$. The simplest action would involve updating sequentially and individually each element $M_t$ of $\va{M}$, for $\forall t \in [0,(2L-1)N]$, conditioned on the state of the rest $\{M_s : s \neq t\}$ of the elements (here $t$ stands for \textit{target} particle, and $s$ for \textit{surroundings}). This scheme satisfies local detailed balance,\cite{Manousiouthakis99, Faizi2020} so
\begin{align}
    \prob{M_t'}{\{M_s\}} \cdot W(M_t,\{M_s\}) = \prob{M_t}{\{M_s\}} \cdot W(M_t',\{M_j\})
    ,
\end{align}
which implies
\begin{align}
\label{eq:cond_prob}
    \prob{M_t'}{\{M_s\}} = \frac{W(M_t',\{M_s\})}{\sum_{M_t} W(M_t,\{M_j\})} .
\end{align}
Since $M_t$ belongs to only two clusters, all the terms in the product but two will cancel out in the numerator and denominator of Eq. \eqref{eq:cond_prob}. Therefore, the conditional probability of updating $M_t$ is
\begin{align}
    \label{eq:cond_prob_matrix_elements}
    \prob{M_t}{\{M_s\}} = \frac{\mel**{M_{s\sub{1}},M_{s\sub{2}}}{}{M_{s\sub{3}},M_t}\mel**{M_t,M_{s\sub{4}}}{}{M_{s\sub{5}},M_{s\sub{6}}}}{\sum_{M} \mel**{M_{s\sub{1}},M_{s\sub{2}}}{}{M_{s\sub{3}},M}\mel**{M,M_{s\sub{4}}}{}{M_{s\sub{5}},M_{s\sub{6}}}},
\end{align}
where all the possible conditional probabilities for all possible types of two clusters interactions can be pre-computed and stored. That way, the sampling scheme proceeds by selecting a target particle, defining its clusters, and then assigning a new value, drawn according to the conditional probabilities. This defines the main idea behind the Gibbs sampling procedure used throughout the PIGS simulations.

\subsection{\label{sec:cluster_loop_moves}Cluster-Loop Moves}
Although the sequential/individual particle updating procedure described in the section \ref{sec:gibbs_samp} follows local detailed balance, it suffers from ergodicity issues. This is related to the symmetry of the Hamiltonian with respect to the parity of the total angular momentum. We can define 
\begin{align}
    \oper{\pi}\sub{\mathcal{L}} = \frac{\1 - e^{\i \pi \L}}{2} ,
\end{align}
with eigenvalues
\begin{equation}
    \frac{1-(-1)^{\mathcal{L}(\m{}{})}}{2} = 
    \left\{
        \begin{aligned}        
            0 & \text{, for $\mathcal{L}(\m{}{})$ even,}\\
            1 & \text{, for $\mathcal{L}(\m{}{})$ odd,}  			
        \end{aligned}
        \right.
\end{equation}
where $\mathcal{L}(\m{}{}) \equiv \sum_{i=1}^N m_i$. It can be shown (see Supplementary Material) that 
\begin{align}
    \comm{\oper{\pi}\sub{\mathcal{L}}}{\H} = 0 \implies \comm{\oper{\pi}\sub{\mathcal{L}}}{\oper{\rho}(\tau)} = 0
    ,
\end{align}
such that the propagated states will preserve the parity of the total angular momentum. Thus, the chosen trial state of Eq. \eqref{eq:trial_state} bounds the propagated states to even parity.

\begin{figure}[ht]
  \centering 
  \vspace{0.3cm}
  \includegraphics[width = 0.6\columnwidth]{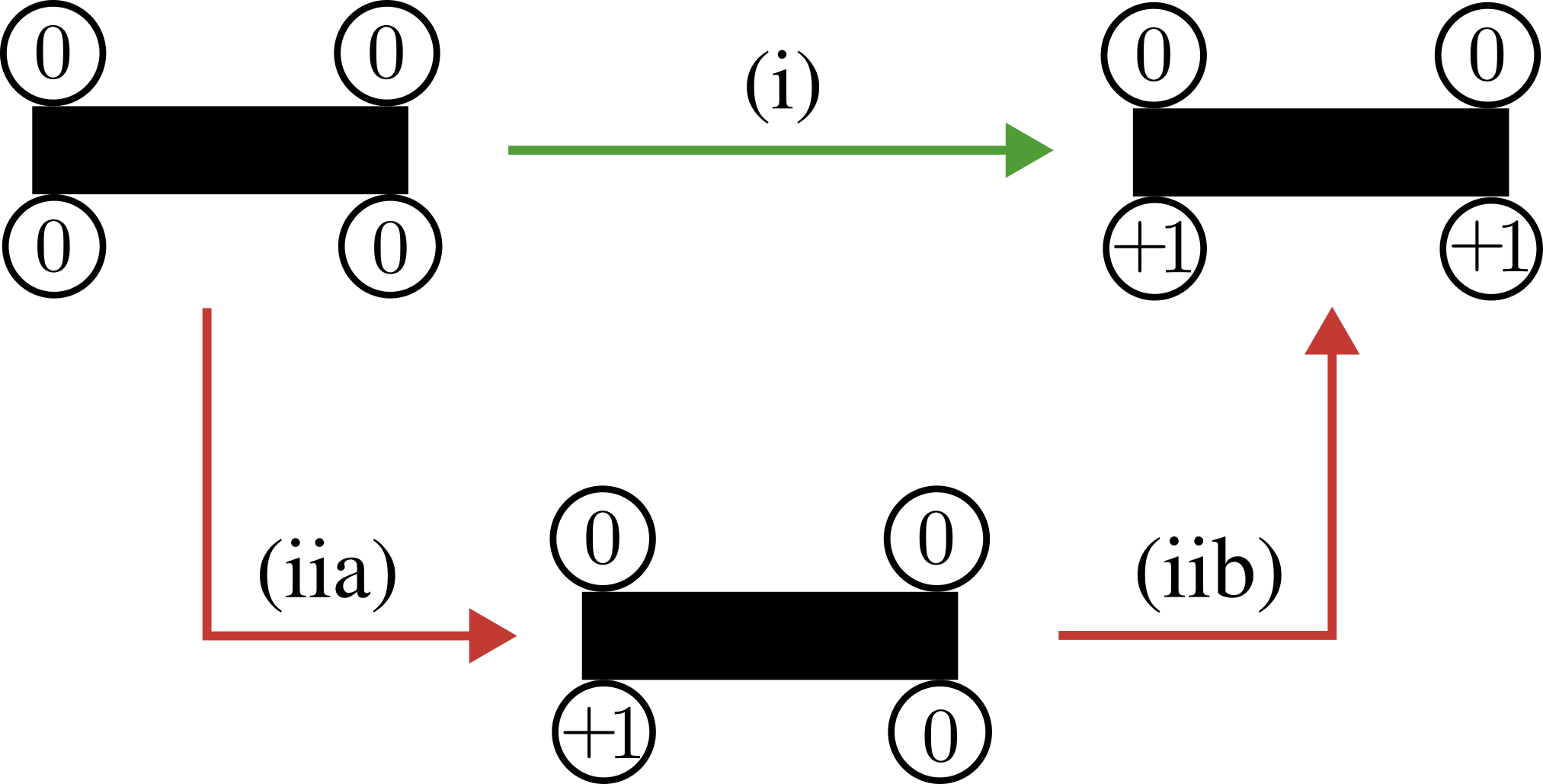}
  \caption{Graphical representation of the process of updating a cluster by changing the particles in two different manners: (i) in pairs; and (ii) individually and sequentially.}
  \label{fig:transitions_symmetry}
\end{figure}

As a consequence, individual particle updates considerably limit the sampling space. This can be seen when considering, for example, the transition of the cluster $\mel**{0,0}{}{0,0}$ to $\mel**{0,0}{}{+1,+1}$ in two different processes: (i) by updating the particles $\m{i}{p+1}$ and $\m{i+1}{p+1}$ altogether, and (ii) by updating first $\m{i}{p+1}$ and then $\m{i+1}{p+1}$, as depicted in Fig. \ref{fig:transitions_symmetry}. The transition probabilities are defined as
\begin{align}
\label{eq:transitions_symmetry_P_(i)}
    \mathrm{P}_{(i)} = \frac{\mel**{0,0}{}{1,1}}{\sum_{m,l} \mel**{0,0}{}{m,l}}
    ,
\end{align}
and
\begin{align}
\label{eq:transitions_symmetry_P_(ii)}
    \mathrm{P}_{(ii)} = \frac{\mel**{0,0}{}{1,0}}{\sum_{m} \mel**{0,0}{}{m,0}} \frac{\mel**{0,0}{}{1,1}}{\sum_{l} \mel**{0,0}{}{1,l}}
    .
\end{align}

In order to make a quantitative comparison we set $L=26 \implies \tau \approx 0.38$ so that $\mathrm{P}_{(i)} \approx 0.133$ while $\mathrm{P}_{(ii)} \approx 10^{-18} \sim 0$. This happens because the cluster $\mel**{0,0}{}{+1,0}$ in the numerator of Eq. \eqref{eq:transitions_symmetry_P_(ii)} violates the conservation of the parity of the total angular momentum, and therefore, has negligible weight contribution ($\approx 10^{-18} \sim 0$), even though the clusters $\mel**{0,0}{}{0,0}$ and $\mel**{0,0}{}{+1,1}$ have relevant weights of $\approx 1.08$ and $\approx 0.25$ respectively. In fact, one can easily check that the stochastic matrix associated with   process (i) is a primitive matrix, and thus, by the Perron-Frobenius theorem, the correspondent Markov chain converges to its stationary point.\cite{meyer2000matrix} The same cannot be said regarding the stochastic matrix associated with process (ii).

\begin{figure}[htb]
    \centering
    \subfloat[]{\includegraphics[width = 0.45\columnwidth, trim={1.0cm 0 5.57cm 0},clip]{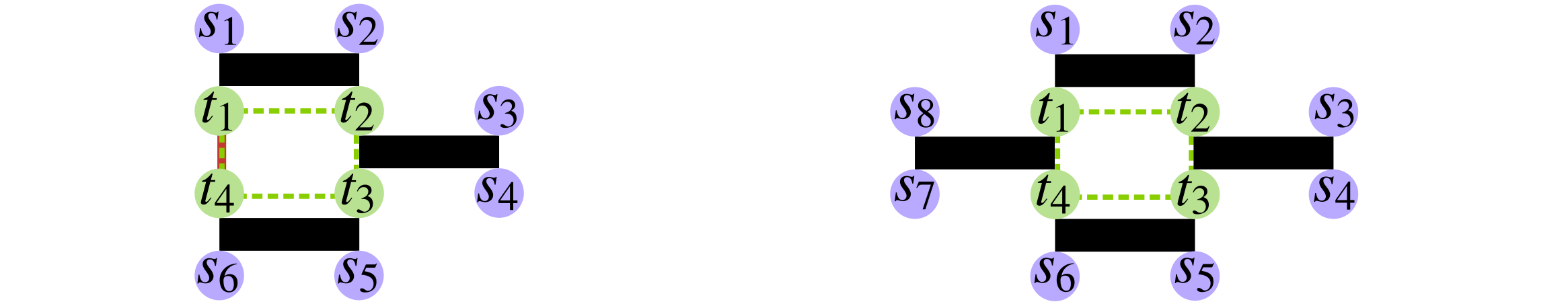}
    \label{fig:PIGS_loops_edge}} 
    \qquad
    \subfloat[]{\includegraphics[width = 0.45\columnwidth, trim={5.57cm 0 1.0cm 0},clip]{PIGS_loops.png}
    \label{fig:PIGS_loops_middle}}
    \caption{Representation of the two simplest possible closed loops (in light green color) connecting pairs of particles that share the same cluster in the PIGS grid.}
    \label{fig:PIGS_loops}
\end{figure}

Consequently, the updating process comes with a condition that strongly affects ergodicity: particles should be selected in such a way that all the clusters involved contain at least one pair of targeted particles. One can achieve this by constructing closed loops connecting pairs of particles that share the same cluster. The generalization of this process leads to the Directed Loop updating scheme\cite{DirLoop1,DirLoop2,DirLoop3}. Here, we will stick to the two simplest possible cases, depicted in Fig. \ref{fig:PIGS_loops}. 
The loop of type \ref{fig:PIGS_loops_edge} contains three particles ($t_4 \equiv t_1$) to be updated conditioned to six other ones in the three clusters, and can be placed on the left or right side edges of the PIGS grid. Also, the loop of type \ref{fig:PIGS_loops_middle} contains four particles to be updated conditioned to eight other ones in the four clusters, and can be placed anywhere in the middle of the PIGS grid. The conditional probabilities for both loops are as follow
\begin{align}
    \label{eq:cond_prob_3cluster}
    \prob{\{t_i\}_3}{\{s_j\}_6} =  
    \frac{
        \mel**{s\sub{1},s\sub{2}}{}{t\sub{1},t\sub{2}}\mel**{t\sub{2},s\sub{3}}{}{t\sub{3},s\sub{4}}\mel**{t\sub{1},t\sub{3}}{}{s\sub{6},s\sub{5}}
    }{
        \sum_{m,l,r} 
        \mel**{s\sub{1},s\sub{2}}{}{m,l}\mel**{l,s\sub{3}}{}{r,s\sub{4}}\mel**{m,r}{}{s\sub{6},s\sub{5}}
    },
\end{align}
\begin{align}
    \label{eq:cond_prob_4cluster}
    &\prob{\{t_{i}\}_4}{\{s_j\}_8} = 
    \nn
    &~~~~~~~~
    \frac{
        \mel**{s\sub{1},s\sub{2}}{}{t\sub{1},t\sub{2}}\mel**{t\sub{2},s\sub{3}}{}{t\sub{3},s\sub{4}}\mel**{t\sub{4},t\sub{3}}{}{s\sub{6},s\sub{5}}\mel**{s\sub{8},t\sub{1}}{}{s\sub{7},t\sub{4}}
    }{
        \sum_{m,l,r,s} 
        \mel**{s\sub{1},s\sub{2}}{}{m,l}\mel**{l,s\sub{3}}{}{r,s\sub{4}}\mel**{s,r}{}{s\sub{6},s\sub{5}}\mel**{s\sub{8},m}{}{s\sub{7},s}
    },
\end{align}
and can be calculated, stored, and easily accessed during the sampling process. Then, each MC step now consists of selecting and successively updating cluster-loops until the entire PIGS grid has been updated. 

\section{Results and Discussion}
\label{sec:Results}

\subsection{Convergence of the Trotter expansion for $N=3$}
\label{sec:tau_convergency}

As mentioned in Sec. \ref{sec:PIGS_planar_rotors}, the approximation of the density matrix $\oper{\rho}(\tau)$ as a product of $2$-body density matrices, as in Eq. \eqref{eq:dens_matrix_tau_bond_H}, bears an error associated with the Trotter decomposition, that vanishes in the limit of $\tau \rightarrow \infty$. This unfeasible limit, along with the associated error, adds a caveat in the choice of a finite $\tau$ parameter on a practical calculation of physical observables of Eq. \eqref{eq:expectation_PIGS_trotterized}. So, in order to reach a balance between mitigating both the approximation error and the computational complexity, a $\tau$ convergency analysis is required.

For a small system of $N=3$ planar rotors with $\mmax=5$, the expectation values of the kinetic, potential and total energy of the ground state are directly calculated by computing the sums of \eqref{eq:expectation_PIGS_trotterized} using NMM for the estimators of Eqs. \eqref{eq:kinetic_estimator} and \eqref{eq:potential_estimator}. Here, both the BH decomposition and the SP decomposition are considered. The exact energies calculated using ED are obtained for comparison.
\begin{figure*}
\includegraphics[width = 1\textwidth]{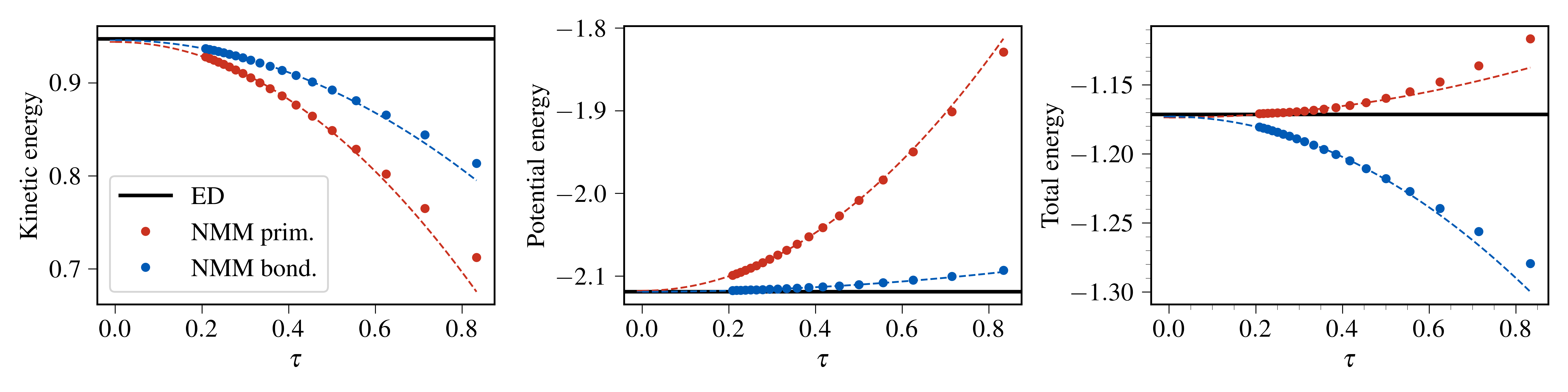}
\caption[]{\label{fig:NMM_vs_ED}Kinetic $\expval{\K{}}$, potential $\expval{\V{}}$, and total energy $\expval{\K{}}+\expval{\V{}}$ as a function of $\tau$ for both the SP Trotter and BH decompositions. The data points for each $\tau$ value were calculated using NMM. ED results are shown as a solid line. The dashed lines are the extrapolation of the NMM data fitted with a polynomial of quadratic order.}
\end{figure*}
The results are shown in Fig. \ref{fig:NMM_vs_ED}, from which we can note the expected behavior for both decompositions in the limit of small $\tau$ ($\tau \sim< 0.5$) where the error is $\mathscr{O}\left(\tau^2\right)$, and the data points converge to the exact ED values when extrapolated using a quadratic fitting function.\cite{suzuki1985, fye1986} We can also see that the BH decomposition outperforms the SP decomposition when calculating the kinetic energy and potential energy estimators for the same values of $\tau$.

The apparent faster convergency of the SP over the BH decomposition when calculating the total energy is due cancellation of errors. This is because the total energy is determined by adding both $\expval{K}$ and $\expval{V}$, calculated separately. Given that the individual accuracy on both $\expval{K}$ and $\expval{V}$ is preferred and required for the calculation of other estimators (as in Eq. \ref{eq:dKdg_estimator}), the BH decomposition will then be chosen to approximate the system's propagator from now on. A chosen range of $\tau \in [0.2, 0.4] \implies L \in [26,50]$ is practical for simulations of larger systems with both acceptable errors and computational costs\footnote{A crude complexity analysis of the algorithm used gives $\mathscr{O}(10^3)$ seconds for each MC simulation of $10^5$ steps for a system of $\mathscr{O}(10^2)$ planar rotors.}.

\subsection{Angular Momentum Properties for $N=150$}
\label{sec:ang_mom_properties}
\begin{figure}[ht]
    \centering
    \includegraphics[width = 0.9\columnwidth, trim={0 0 0 0},clip]{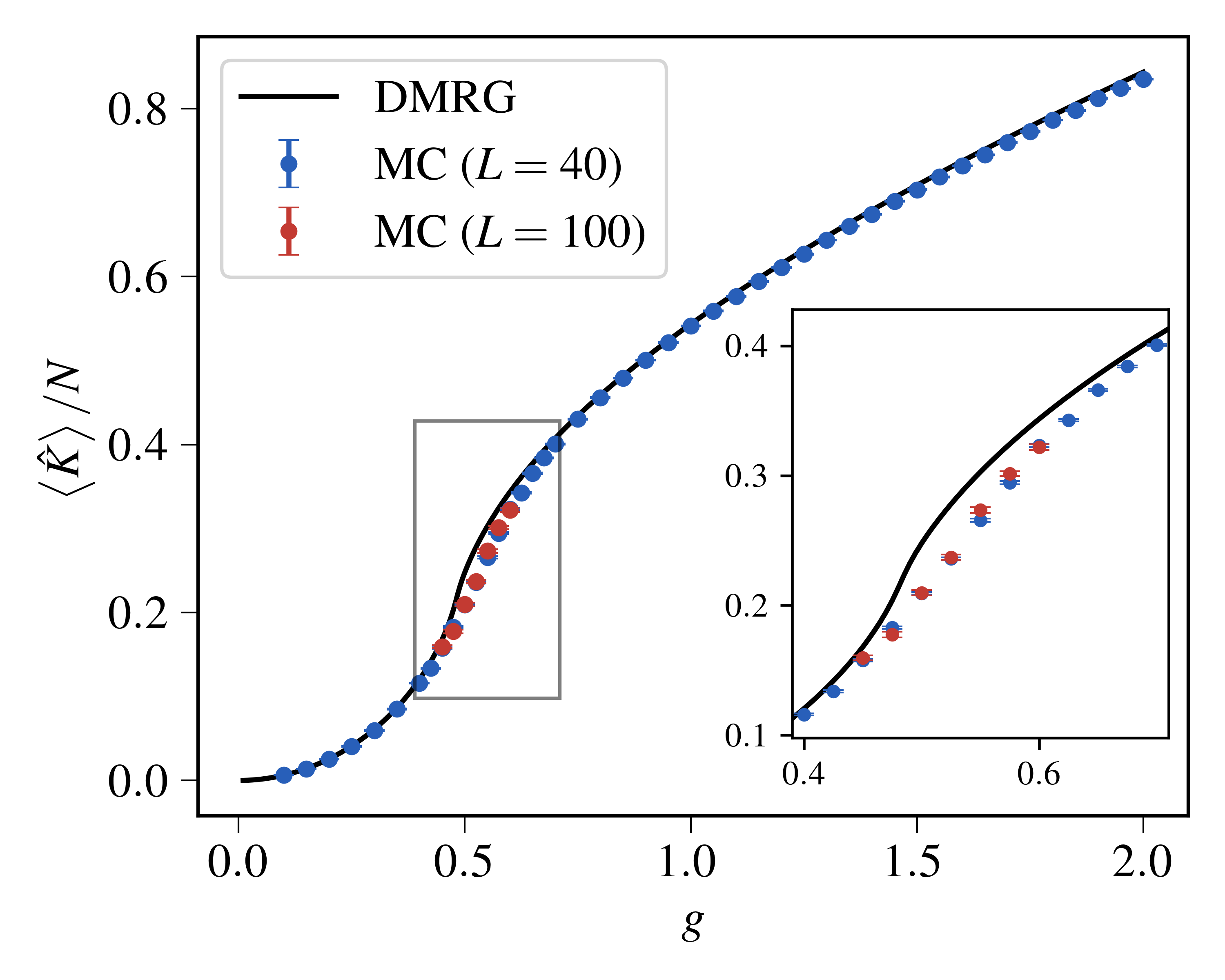}
    \caption[]{\label{fig:KE_vs_g}Expectation value of the ground state kinetic energy $\expval{\K{}}$ of the system as a function of the interaction strength $g$ for $N=150$ planar rotors with $\mmax=5$. The MC results were calculated for $L=40$ and $L=100$ beads (around the critical region), both using $10^5$ MC steps. The DMRG result is shown for comparison.}
\end{figure}

\begin{figure}[ht] 
    \centering
    \includegraphics[width = 0.9\columnwidth, trim={0 0 0 0},clip]{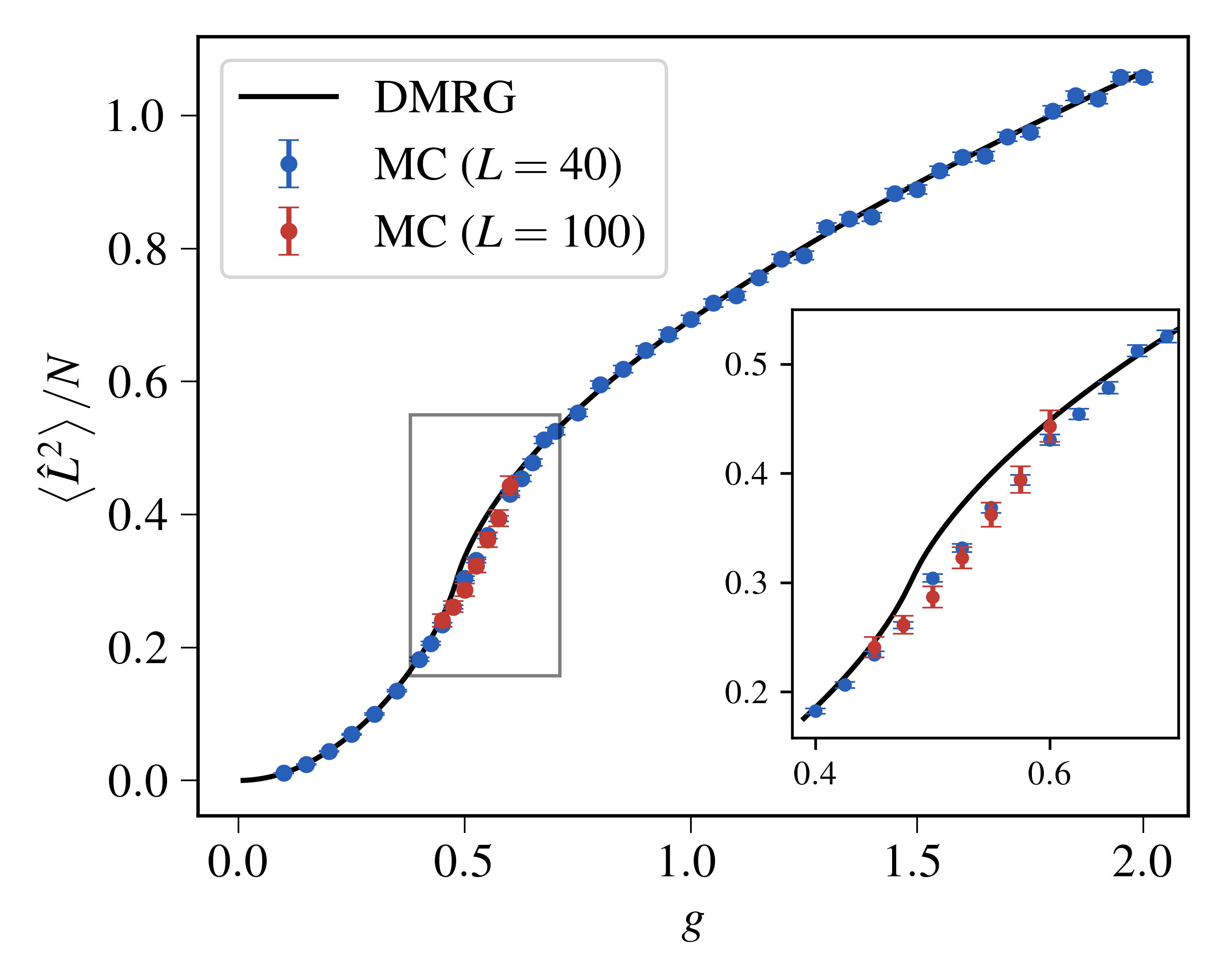}
    \caption[]{\label{fig:L2_vs_g}Expectation value of the ground state total angular momentum squared $\expval{\L^{2}}$ as a function of the interaction strength $g$ for $N=150$ planar rotors with $\mmax=5$. The MC results were calculated for $L=40$ and $L=100$ beads (around the critical region), both using $10^5$ MC steps. The DMRG result is shown for comparison.}
\end{figure}

For a large system consisting of $N=150$ rotors, we present in  Figs. \ref{fig:KE_vs_g} and \ref{fig:L2_vs_g} the kinetic energy and the total angular momentum squared obtained from Eqs. \eqref{eq:kinetic_estimator} and \eqref{eq:L2_estimator} respectively for a series of $g$ values.
The MC results for both estimators are compared with the DMRG calculations. Details of the DMRG implementation can be found in Ref. \onlinecite{Serwatka2024} where the ITensor package\cite{fishman2022itensor} is used for computations. 


It has previously been established\cite{Serwatka2024,Zhang2024,Moed2024} that this system undergoes a QPT for some critical dipole-dipole interaction strength $g_c \sim 0.485$. Here, the change in slope of the curves in Figs. \ref{fig:KE_vs_g} and \ref{fig:L2_vs_g} is an indcator of this criticality. Apart from the critical region, highlighted in the insets for $0.4 \le g \le 0.7$, the MC method agrees very well with the DMRG benchmark. The discrepency observed around the phase transition could be a symptom of the critical slowing down, which can be mostly attributed to the fact that the cluster-loop sampling scheme relies on local moves only, even though rotors are not updated individually.\cite{WOLFF199093, Bonati2018} 
Another reason for the observed mismatch could be the finite size effect in the imaginary time dimension and the Trotter error. Considering that the original $1-d$ quantum problem is mapped into a $(1+1)-d$ classical one, the same type of effects observed for finite size lattices such as the $2-d$ Ising Model,\cite{binder1981finite} might be observed. Besides the finite size in the spatial dimension (that also affects the DMRG results), we also have a finite $\beta$ imaginary time dimension. Moreover, the Trotter error might exhibit a non trivial dependence on the interaction strength $g$. 

The similarity between the results, especially away from the critical point, for $L=40$ (blue points) and $L=100$ (red points) appears to indicate that the BH decomposition of the propagator in the angular momentum basis is numerically impervious to $\tau$ errors. However, it could also be possible that a much higher number of beads would result in a gradual convergence towards the DMRG results and highlight the influence of such imaginary time finite-size effect related errors.
This type of problem was also observed for the pair product approximation in the angular position basis $\{\ket{\phi_i}\}$ in Ref. \onlinecite{Moed2024}. 


A further consideration to be made is that although the definition of the order parameter given in terms of the angular orientation of the rotors in Refs. \onlinecite{Serwatka2024} and \onlinecite{Moed2024} can be computed in a rather straightforward manner in the $\{\ket{\phi_i}\}$ position basis, things are not as simple here. This difficulty comes from the fact that observables in the $\{\ket{m_i}\}$ basis are represented by off-diagonal operators with long range action --- e.g. higher moments of the Polarization estimator --- cannot be decomposed in the form of Eq. \eqref{eq:bond-O_odd_even_def}, leading to a different definition for the estimators that implies extra computational cost, as mentioned in Sec. \ref{sec:PIGS_Estimators}.

\begin{figure}[ht]
    \centering
    \includegraphics[width = 0.9\columnwidth, trim={0 0 0 0},clip]{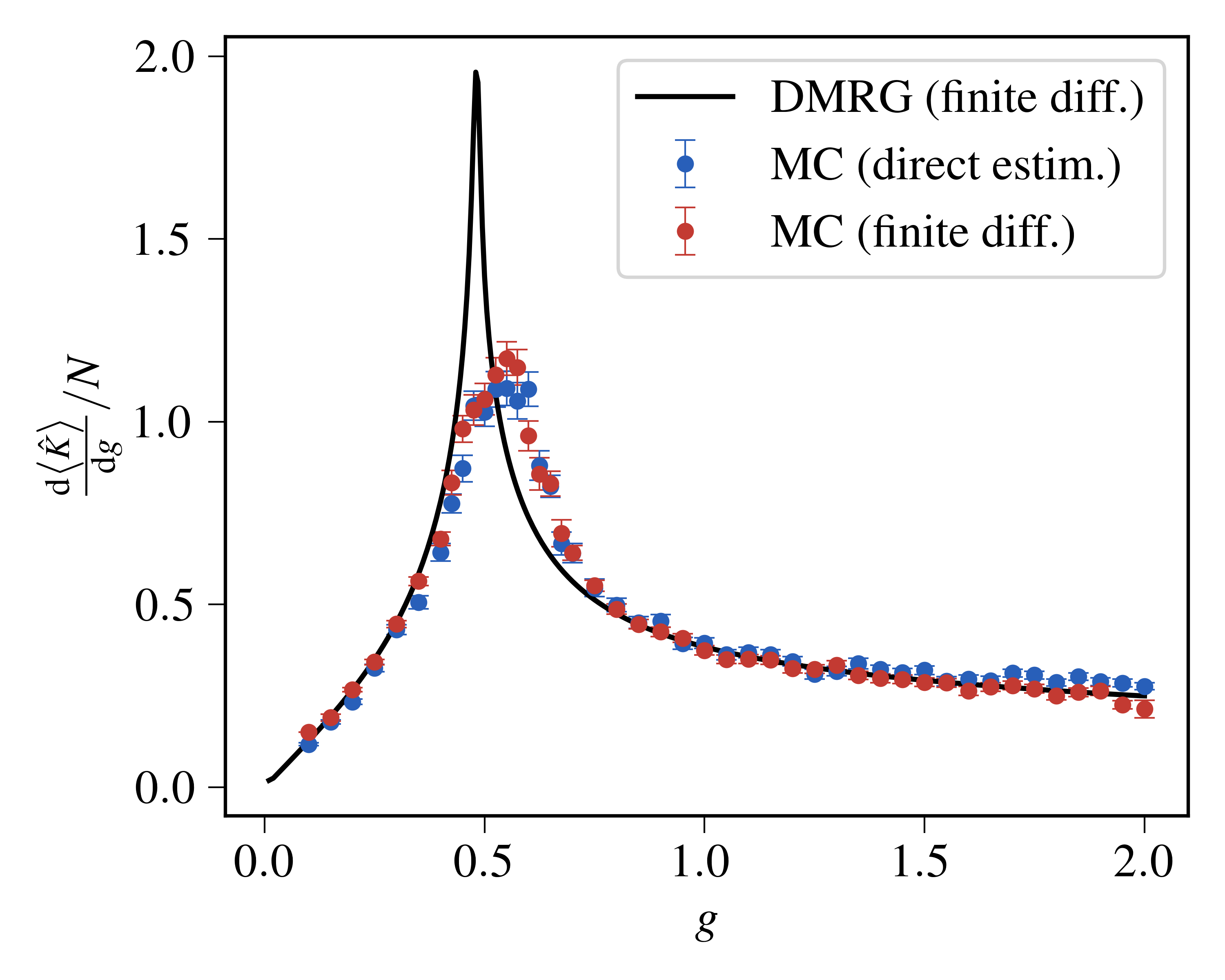}
    \caption[]{First derivative of the expectation value for the ground state kinetic energy $\expval{\K{}}$ with respect to the interaction strength $g$ for $N=150$ planar rotors with $\mmax=5$.} 
    \label{fig:dKdg_vs_g}
\end{figure}

However, the QPT can still be observed from the analysis of the observables proposed here despite the absence of an actual definition and/or computation of a suitable order parameter in the angular momentum basis. The inflection point on the curves $\expval{\K{}}/N$ and $\expval{\L^2}/N$ around $g_c$ is clear evidence of that. This manifestation of criticality can be highlighted by computing the first derivatives of both quantities with respect to $g$. The first derivative $\expval{\K{}}$ with respect to $g$ was calculated using both a direct MC estimator from Eq. \eqref{eq:dKdg_estimator} and also using the finite difference of the MC and DMRG curves from Fig. \ref{fig:KE_vs_g}. The results are shown in Fig. \ref{fig:dKdg_vs_g}.
The peak of the curves in the critical region is an indicator of the QPT, where the singularity would be reached in the limit $N\rightarrow \infty$. The critical value of the coupling strength is estimated at $g_c \sim 0.485$ for the DMRG result whereas $g_c \sim 0.525$ for MC results and this shift observed is again due to the critical slowdown for the MC simulations, along with $\tau$ and $\beta$ errors.

The agreement between the curve for the MC direct estimator from Eq. \eqref{eq:dKdg_estimator} and the results obtained by taking finite differences of $\expval{\K{}}(g)$ (computed using MC), shows the success of the proposed method in directly estimating the structural property $\dv*{\expval{\K{}}}{g}$, which we use as our QPT indicator.. Here it is important to note that the same process could have been done for $\dv*{\expval{\L{}^2}}{g}$. Nonetheless, from Eq. \eqref{eq:L2_estimator} we have
\begin{align}
    \mathcal{L}^2 = \left(\sum_{i=1}^N \m{i}{L+1}\right)^2 = \mathcal{K} + \sum_{i=1}^N \sum_{j\neq i} \m{i}{L+1}\m{j}{L+1},
\end{align}
and to illustrate this idea, we can simply use the DMRG results to compare the first derivatives of $\expval{\K{}}$, $\expval{\L^2}$, and $\expval{\L^2} - \expval{\K{}}$ with respect to $g$, calculated using the finite difference method.
\begin{figure}[ht]
    \centering
    \includegraphics[width = 0.9\columnwidth, trim={0 0 0 0},clip]{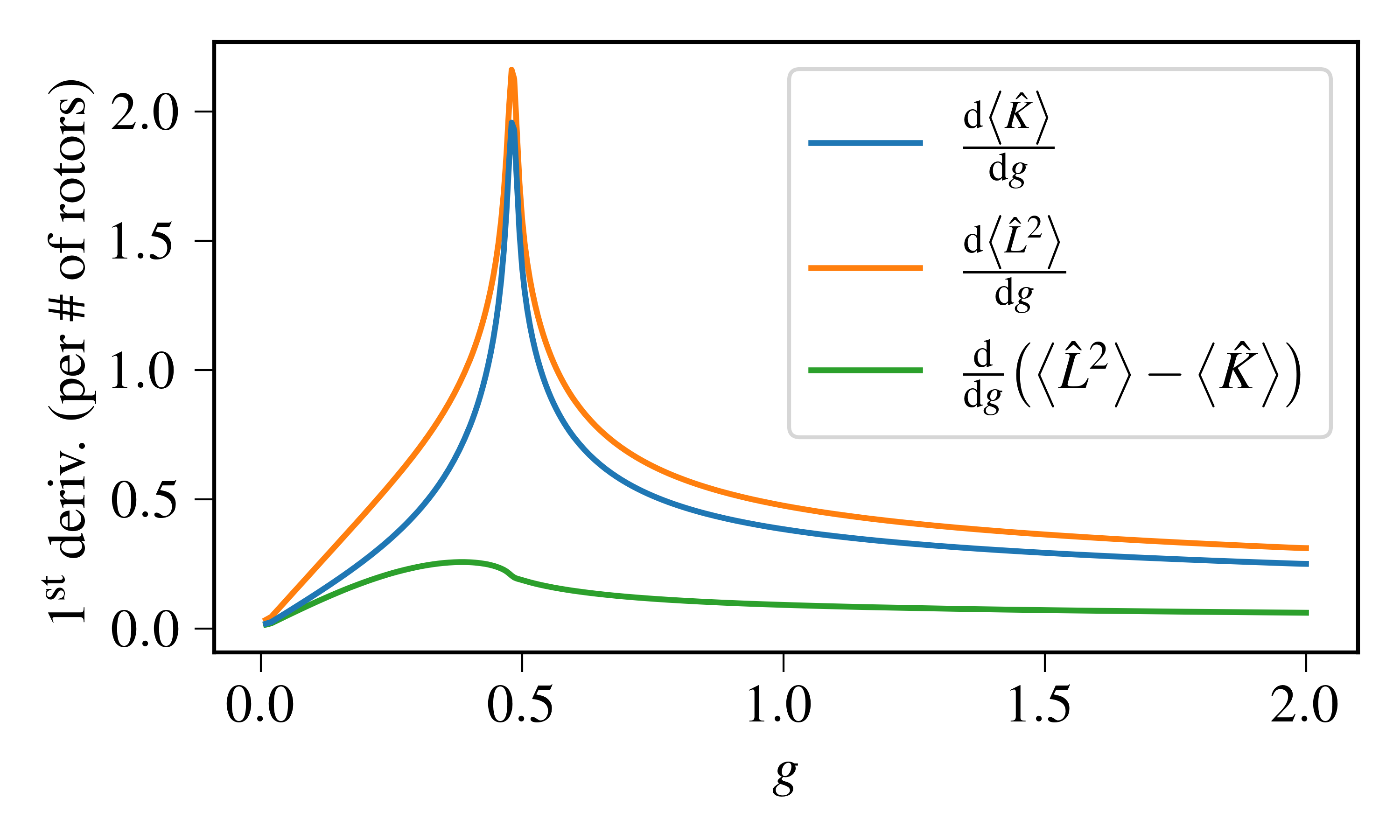}
    \caption[]{First derivatives of the expectation values $\expval{\K{}}$, $\expval{\L^2}$, and $\expval{\L^2} - \expval{\K{}}$ with respect to the interaction strength $g$ for $N=150$ planar rotors with $\mmax=5$.} 
    \label{fig:first_derivs_DMRG}
\end{figure}
From Fig. \ref{fig:first_derivs_DMRG} it becomes obvious that most of the singularity in $\expval{\L^2}$ comes from the $\expval{\K{}}$ observable.

\subsection{Ordered and Disordered phases}

A direct consequence of the invariance with respect to a collective rotation of $\pi$ of all dipoles, the $\Zmath_2$ symmetry, is the appearance of a twofold degenerate ground state in the ordered phase, represented in the $\{\ket{\phi}\}$ basis by
\begin{align}
    \ket{\Psi}^{\text{ord}}_{\text{gs}} =  \ket{0,0,\dots,0} + \ket{\pi,\pi,\dots,\pi} 
\end{align}
up to a normalization factor.\cite{Serwatka2024} 
In the $\{\ket{m}\}$ basis we have
\begin{align} 
    \label{eq:ordered_state}
    \ket{\Psi}^{\text{ord}}_{\text{gs}} 
    =&\sum_{\m{}{}} \ketbra{\m{}{}}\left( \ket{0,0,\dots,0} + \ket{\pi,\pi,\dots,\pi} \right)
    \nn
    =& \sum_{\m{1}{}}\sum_{\m{2}{}} \dots \sum_{\m{N}{}} C_{\mathcal{L}({\m{}{}})} \ket{\m{1}{},\m{2}{},\dots,\m{N}{}}
    ,
\end{align}
for $C_{\mathcal{L}({\m{}{}})} \equiv 1 + e^{\i \pi \mathcal{L}({\m{}{}})}$, where  $\braket{m_i}{\phi_i} = e^{\i\pi m_i \phi_i}$ was used.\cite{Kastrup_2006} From Eq. \eqref{eq:ordered_state} it is evident that the ordered state in the angular momentum basis is a superposition state with equal probability amplitude for all states in which the total angular momentum has even parity.

As mentioned in Sec. \ref{sec:cluster_loop_moves}, the absence of states with odd parity is expected, and now the uniform distribution of single particle states that lead to an even parity of total angular momentum emerges as a feature of the ordered phase for $g \ge g_c$. This becomes evident when we analyze the distribution of the total angular momentum observable $\expval{\L}/N$ for various interaction strengths $g$.
\begin{figure*}[htp]
\includegraphics[width = 1\textwidth]{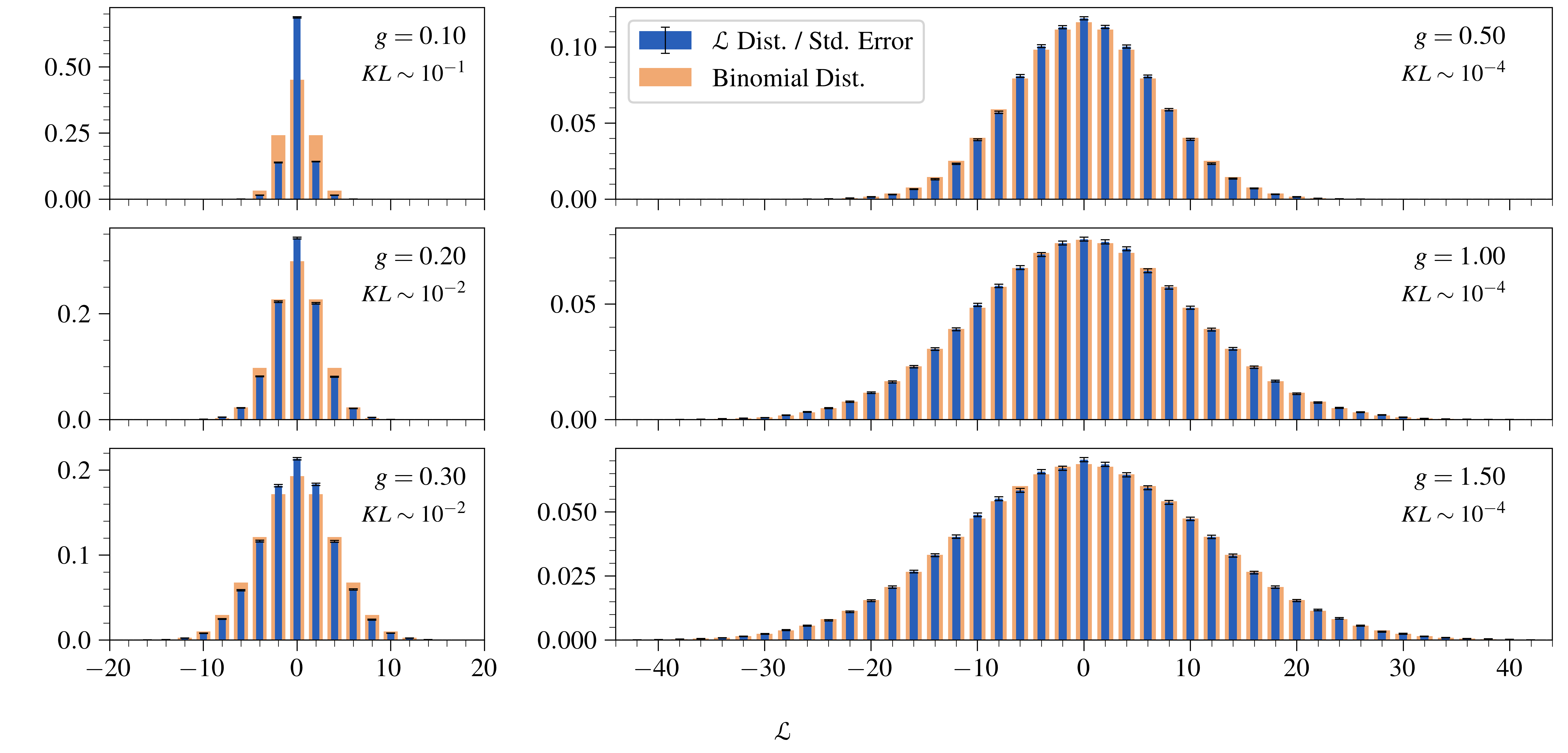}
\caption{\label{fig:histograms_Ltotal} Histograms of the values for total angular momentum $\L{}$ observable during the MC simulations for various dipole-dipole interaction strengths $g$. The orange bars are the binomial distributions with variance equivalent to $\expval{\L^2}/N$ for the respective value of $g$. The relative entropy\cite{Shannon48} between both distributions was also calculated, where $D_{KL}=\sum_i p_i \ln{p_i/q_i}$ for $p_i$ and $q_i$ the $\mathcal{L}$ and the binomial distributions respectively.}
\end{figure*}
From Fig. \ref{fig:histograms_Ltotal} we can see that the probability for total angular momentum with odd parity is zero, and also that the distribution of the even total angular momenta (blue bars) approaches a binomial (normal) distribution (red curves) with mean $\expval{\L}=0$ and variance $\expval{\L^2}/N$, as $g$ approaches $g_c$. For $g>g_c$ the histograms then match the normal distribution.
The relative entropy\cite{Shannon48} or Kullback–Leibler divergence (KL)\cite{kullback1951information} is used here to measure the statistical distance between the distributions. We note that for $g$ values greater than $g_c$, the KL divergence is around  2 orders of magnitude smaller than for $g<g_c$. This is strong evidence that a drop in  KL magnitude is an indicator of the QPT.
This can be translated to the problem of restricted compositions of numbers from number theory \cite{andrews1998theory, eger2013restricted} where in the ordered phase, all the compositions of $\mathcal{L}({\m{}{}})$ are equally probable. In this scenario, the counting of the total number of compositions of $\mathcal{L}({\m{}{}})$, i.e. the total number of states of type $\ket{\m{}{}}$ such that $\sum_i \m{}{i} = \mathcal{L}$ even, is related to the extended binomial coefficients,\cite{eger2013restricted} which may explain the coincidence between the distribution of $\mathcal{L}$ and a normal distribution.
The full connection between the distribution of total angular momentum and the problem of restricted compositions will be the subject of future work.

\section{Concluding remarks}
\label{sec:Conclusions}

We introduced a PIMC approach in the angular momentum representation. A system consisting of identical planar rotors with dipole-dipole interactions was used as an illustrative example.
The formulation of the partition function as discrete path sums on a high rank tensor with finite state probabilities renders the problem suitable for Gibbs sampling Monte Carlo updates.
The choice of factorization  based on the BH decomposition together with the grouping of commuting terms (\textit{odd} and \textit{even} terms) for n.n.i. proved to be not only convenient, but also successful in approximating the system's propagator up to a $\mathscr{O}\left(\tau^2\right)$ error as seen from the $\tau$ convergence analysis of the ground state energy estimators. 

The proposed direct calculation of angular momentum properties was successful and efficient, with excellent agreement (away from the QPT) between the PIGS MC and benchmark DMRG results. Both the kinetic energy and total angular momentum squared were obtained from the MC estimators, and despite the absence of a suitable direct order parameter estimator, the derivative of the kinetic energy with respect to the interaction strength worked as good phase transition indicator. It is important to note that the critical slowing down observed in the MC results could come from the fact that the \textit{cluster-loop} updating scheme in its simplest form relies on local moves only.

Our theoretical analysis has revealed the role played by the parity symmetry of the total angular momentum. As a result, the \textit{cluster-loop} moves emerged as the inherent {\it de facto} requirement for the updating scheme by effectively guaranteeing the non-limitation of the sampling space. Furthermore, other implications of symmetry are (i) the even parity of the ground state and, (ii) the  uniform distribution of individual angular momentum states in the ordered phase, subject to the even parity constraint of total angular momentum. We have presented strong evidence that support the relation of the latter to the problem of restricted compositions, but further research is needed to understand the depth of such a connection. 

In order to mitigate the critical slowing down problem, updating schemes based on non-local moves are good candidates, and here the Directed Loop algorithm\cite{DirLoop1, DirLoop2, DirLoop3} appears as the natural generalization of the present \textit{cluster-loop} updating scheme. Extension to more complex lattices with long range interactions in 2-D or 3-D will pose challenges in the derivation of  MC estimators, especially when dealing with non-local observables. Hence, new MC methods could be combined and applied as auxiliary techniques. Promising candidates are the Stochastic Series Expansion\cite{Sandvik1991,Sandvik_1992} and the Permutation Matrix Representation\cite{Gupta_2020} QMC methods, both successfully applied to a variety of discrete systems,\cite{Zyubin2004, babakhani2025, ezzell2025,kalev2025feynman} with the potential to be efficiently adapted to many-body confined rotor systems.

\section*{Supplementary Material}
See the supplementary material for additional results and derivations from Sec. \ref{sec:Theory} that were omitted from the main text for the sake of brevity. Details of the MC sampling are also provided.

\begin{acknowledgments}
This research was supported by the Natural Sciences and Engineering Research Council (NSERC) of Canada (RGPIN-03725-2022), the Ontario Ministry of Research and Innovation (MRI), the Canada Research Chair program (950-231024), the Digital Research Alliance of Canada, and the Canada Foundation for Innovation (CFI) (project No. 35232).
\end{acknowledgments}

\section*{Data Availability Statement}

The data that support the findings of this study are available
from the corresponding author upon reasonable request.

\bibliography{references}

\pagebreak
\clearpage
\onecolumngrid

{
\sffamily
\bfseries
\linespread{1.5}
\Large 
\noindent
Supplementary Material for: Path Integral Monte Carlo in the Angular Momentum Basis for a Chain of Planar Rotors
}

\setcounter{equation}{0}
\setcounter{figure}{0}
\setcounter{table}{0}
\setcounter{page}{1}
\setcounter{section}{0}
\makeatletter
\renewcommand{\theequation}{S\arabic{equation}}
\renewcommand{\thefigure}{S\arabic{figure}}
\renewcommand{\bibnumfmt}[1]{[S#1]}
\renewcommand{\citenumfont}[1]{S#1}

\section{Additional derivations}
\subsection{Commutation between $\oper{\pi}\sub{\mathcal{L}}$ and $\H$}
\label{sec:commutation_parity_H}

Starting from the definitions of the Hamiltonian of the system
\begin{align}
    \label{eq:hamiltonian_op_quantum_system_SUP}
    \H = \sum_{i=1}^N \K{i} - g \sum_{\expval{ij}} \V{ij},
\end{align}
and the operator for the parity of the total angular momentum
\begin{align}
    \label{eq:definition_parity_op}
    \oper{\pi}\sub{\mathcal{L}} = \frac{\1 - e^{\i \pi \L}}{2} ,
\end{align}
we have
\begin{align}
    \comm{\oper{\pi}\sub{\mathcal{L}}}{\H}
    =& \comm{\frac{\1 - e^{\i \pi \L}}{2}}{\H}
    \nn
    =& \comm{\frac{\1 - e^{\i \pi \sum_{k=1}^N \L_{k}}}{2}}{\sum_{i=1}^N \K{i} - g \sum_{\expval{ij}} \V{ij}}
    \nn
    =&\frac{g}{2} \sum_{\expval{ij}}\comm{e^{\i \pi \sum_{k=1}^N \L_{k}}}{\V{ij}}
    \nn
    =&\frac{g}{2} \sum_{\expval{ij}} \prod_{k\neq i,j} \comm{e^{\i \pi \left(\L_{i} + \L_{j}\right)}}{\V{ij}} .
\end{align}
Here,
\begin{align}
    \comm{e^{\i \pi \left(\L_{i} + \L_{j}\right)}}{\V{ij}} =& ~ e^{\i \pi \left(\L_{i} + \L_{j}\right)}\V{ij} - \V{ij}e^{\i \pi \left(\L_{i} + \L_{j}\right)}
    \nn
    \comm{e^{\i \pi \left(\L_{i} + \L_{j}\right)}}{\V{ij}} \ket{m_i,m_j} =& ~ e^{\i \pi \left(\L_{i} + \L_{j}\right)}\V{ij} \ket{m_i,m_j} - \V{ij}e^{\i \pi \left(\L_{i} + \L_{j}\right)}\ket{m_i,m_j}
    \nn
    =& ~ \left( e^{\i \pi \left(\L_{i} + \L_{j}\right) } - e^{\i \pi \left(m_{i} + m_{j}\right)} \right)\V{ij} \ket{m_i,m_j}
    \nn
    =& ~ \left( e^{\i \pi \left(\L_{i} + \L_{j}\right) } - e^{\i \pi \left(m_{i} + m_{j}\right)} \right) 
        \times \nn & \hspace{1cm} \times 
         \left( \frac{3}{4}\ket{m_i+1,m_j+1} + \frac{1}{4}\ket{m_i+1,m_j-1} + \frac{1}{4}\ket{m_i-1,m_j+1} + \frac{3}{4}\ket{m_i-1,m_j-1}\right),
\end{align}
where we can notice that 
\begin{align}
    \left( e^{\i \pi \left(\L_{i} + \L_{j}\right) } - e^{\i \pi \left(m_{i} + m_{j}\right)} \right)  \ket{m_i\pm1,m_j\pm1} 
    =& \left( e^{\i \pi \left(m_i + m_j \pm 2\right) } - e^{\i \pi \left(m_{i} + m_{j}\right)} \right)  \ket{m_i\pm1,m_j\pm1}
    \nn
    =& \left( e^{\pm \i 2\pi } - 1\right)  e^{\i \pi \left(m_{i} + m_{j}\right)}  \ket{m_i\pm1,m_j\pm1}
    \nn
    =& ~ 0 .
\end{align}
Therefore,
\begin{align}
    \comm{e^{\i \pi \left(\L_{i} + \L_{j}\right)}}{\V{ij}} \ket{m_i,m_j} = 0
    \implies 
    \comm{e^{\i \pi \left(\L_{i} + \L_{j}\right)}}{\V{ij}} = 0,
\end{align}
implying that
\begin{align}
    \label{eq:commutation_pi_H}
    \comm{\oper{\pi}\sub{\mathcal{L}}}{\H}
    = 0.
\end{align}

\subsection{Commutation between $\oper{\pi}\sub{\mathcal{L}}$ and $\oper{\rho}(\tau)$}
\label{sec:commutation_parity_propagator}

Starting from the definition of $\oper{\rho}(\tau) \equiv e^{- \tau \H}$ we have
\begin{align}
    \comm{\oper{\pi}\sub{\mathcal{L}}}{\oper{\rho}(\tau)}
    = \comm{\oper{\pi}\sub{\mathcal{L}}}{e^{- \tau \H}}
    = \sum_{n=0}^{\infty} \frac{(-\tau)^n}{n!} \comm{\oper{\pi}\sub{\mathcal{L}}}{\H^n} .
\end{align}
From the commutator we get the recursion relation
\begin{align}
    \comm{\oper{\pi}\sub{\mathcal{L}}}{\H^n}
    =& \oper{\pi}\sub{\mathcal{L}} \H^n - \H^n \oper{\pi}\sub{\mathcal{L}}
    \nn
    =& \oper{\pi}\sub{\mathcal{L}} \H \H^{n-1} - \H^n \oper{\pi}\sub{\mathcal{L}}    
    \nn
    =& \oper{\pi}\sub{\mathcal{L}} \H \H^{n-1} - \H \oper{\pi}\sub{\mathcal{L}} \H^{n-1} + \H \oper{\pi}\sub{\mathcal{L}} \H^{n-1} - \H^n \oper{\pi}\sub{\mathcal{L}}
    \nn
    =& \comm{\oper{\pi}\sub{\mathcal{L}}}{ \H} \H^{n-1} + \H \comm{ \oper{\pi}\sub{\mathcal{L}} }{ \H^{n-1}}
    ,
\end{align}
that gives
\begin{align}
    \comm{\oper{\pi}\sub{\mathcal{L}}}{\H^n}
    = \sum_{p=1}^n  \H^{p-1} \comm{\oper{\pi}\sub{\mathcal{L}}}{ \H} \H^{n-p} 
    = 0
    ,
\end{align}
where in the last step it was used the result from Eq. \eqref{eq:commutation_pi_H}, and hence,
\begin{align}
    \comm{\oper{\pi}\sub{\mathcal{L}}}{\oper{\rho}(\tau)} = 0 .
\end{align}

\subsection{Derivative of $\expval{K}$ with respect to $g$}
\label{sec:dKdg_derivation}

Starting with
\begin{align}		
\label{eq:partition_function_PIMC_trotter_Heven_Hodd_ini_SUP}
    \Z(\beta) 
    \approx \sum_{\{\m{}{p}\}_{2L}} \prod_{p=1}^{2L} \mel*{\m{}{p}}{\oper{\rho}_p(\tau)}{\m{}{p+1}},
\end{align}
for
\begin{equation}
    \label{eq:def_rho_tau_op_SUP}
    \oper{\rho}_p(\tau) =
    \left\{
    \begin{aligned}
        e^{ -\tau \H_{\text{odd}}} & \text{, for $p$ odd,} \\
        e^{ -\tau \H_{\text{even}}} & \text{, for $p$ even,} 			
    \end{aligned}
    \right.
\end{equation}
we have
\begin{align}
  \expval{\K{}} = \frac{1}{\Z}  \sum_{\{\m{}{l}\}_{2L}} \mathcal{K} \prod_{p=1}^{2L} \mel*{\m{}{p}}{\oper{\rho}_p(\tau)}{\m{}{p+1}},
\end{align}
for $\mathcal{K} \equiv \mel{\m{}{L+1}}{\K{}}{\m{}{L+1}} = \left(\m{}{L+1}\right)^2 = \sum_{i=1}^N \left(\m{i}{L+1}\right)^2$. Then,
\begin{align}
    \label{eq:dKdg}
  \dv{\expval{\K{}}}{g} 
    =& \dv{\Z^{-1}}{g} \sum_{\{\m{}{l}\}_{2L}} \mathcal{K} \prod_{p=1}^{2L} \mel*{\m{}{p}}{\oper{\rho}_p(\tau)}{\m{}{p+1}}
    + \frac{1}{\Z} \sum_{\{\m{}{l}\}_{2L}} \mathcal{K} \dv{g} \prod_{p=1}^{2L} \mel*{\m{}{p}}{\oper{\rho}_p(\tau)}{\m{}{p+1}}
    \nn
    =& -\frac{1}{\Z} \dv{\Z}{g} ~ \frac{1}{\Z} \sum_{\{\m{}{l}\}_{2L}} \mathcal{K} \prod_{p=1}^{2L} \mel*{\m{}{p}}{\oper{\rho}_p(\tau)}{\m{}{p+1}}
    + \frac{1}{\Z} \sum_{\{\m{}{l}\}_{2L}} \mathcal{K} \dv{g} \prod_{p=1}^{2L} \mel*{\m{}{p}}{\oper{\rho}_p(\tau)}{\m{}{p+1}}
    \nn
    =& - \expval{\K{}} \frac{1}{\Z} \sum_{\{\m{}{l}\}_{2L}} \dv{g} \prod_{p=1}^{2L} \mel*{\m{}{p}}{\oper{\rho}_p(\tau)}{\m{}{p+1}}
    + \frac{1}{\Z} \sum_{\{\m{}{l}\}_{2L}} \mathcal{K} \dv{g} \prod_{p=1}^{2L} \mel*{\m{}{p}}{\oper{\rho}_p(\tau)}{\m{}{p+1}}
    .
\end{align}
The derivative of the product is
\begin{align}
     \dv{g} \prod_{p=1}^{2L} \mel*{\m{}{p}}{\oper{\rho}_p(\tau)}{\m{}{p+1}}
     =& \sum_{q=1}^{2L} \mel*{\m{}{q}}{\dv{\oper{\rho}_q(\tau)}{g}}{\m{}{q+1}} \prod_{p\neq q} \mel*{\m{}{p}}{\oper{\rho}_p(\tau)}{\m{}{p+1}}
     \nn
     =& \sum_{q=1}^{2L} 
     \frac{\mel*{\m{}{q}}{\dv{\oper{\rho}_q(\tau)}{g}}{\m{}{q+1}}}{\mel*{\m{}{q}}{\oper{\rho}_q(\tau)}{\m{}{q+1}}} 
     \prod_{p=1}^{2L} \mel*{\m{}{p}}{\oper{\rho}_p(\tau)}{\m{}{p+1}}
     ,
\end{align}
where we can write
\begin{equation}
    \dv{\oper{\rho}_q(\tau)}{g} =
    \left\{
    \begin{aligned}
        \dv{g} e^{ -\tau ~\H_{\text{odd}}} & \approx \tau \V{\text{odd}} e^{ -\tau \H_{\text{odd}}}, & \text{ for $q$ odd,} \\
        \dv{g} e^{ -\tau ~ \H_{\text{even}}} & \approx \tau e^{ -\tau \H_{\text{even}}} \V{\text{even}}, & \text{ for $q$ even,} 			
    \end{aligned}
    \right.
\end{equation}
so that
\begin{align}
     \dv{g} \prod_{p=1}^{2L} \mel*{\m{}{p}}{\oper{\rho}_p(\tau)}{\m{}{p+1}}
     =& \tau \sum_{q=1}^{2L} 
     \frac{\mel*{\m{}{q}}{\dv{\oper{\rho}_q(\tau)}{g}}{\m{}{q+1}}}{\mel*{\m{}{q}}{\oper{\rho}_q(\tau)}{\m{}{q+1}}} 
     \prod_{p=1}^{2L} \mel*{\m{}{p}}{\oper{\rho}_p(\tau)}{\m{}{p+1}}
     \nn
     =& \beta \frac{1}{L}\sum_{l=1}^{L} 
     \frac{\mel*{\m{}{2l-1}}{ \V{\text{odd}} e^{ -\tau \H_{\text{odd}}} }{\m{}{2l}}}{\mel*{\m{}{2l-1}}{e^{ -\tau \H_{\text{odd}}}}{\m{}{2l}}} 
     +\frac{\mel*{\m{}{2l}}{ e^{ -\tau \H_{\text{odd}}} \V{\text{even}} }{\m{}{2l+1}}}{\mel*{\m{}{2l}}{e^{ -\tau \H_{\text{even}}}}{\m{}{2l+1}}}
     \prod_{p=1}^{2L} \mel*{\m{}{p}}{\oper{\rho}_p(\tau)}{\m{}{p+1}}
     \nn
     =& \beta \frac{1}{L}\sum_{l=1}^{L} 
        \left( \mathcal{V}_{l,\text{odd}} + \mathcal{V}_{l,\text{even}}\right)
     \prod_{p=1}^{2L} \mel*{\m{}{p}}{\oper{\rho}_p(\tau)}{\m{}{p+1}}
     \nn
     =& \sum_{l=1}^{L} 
     \mathcal{V}_l
     \prod_{p=1}^{2L} \mel*{\m{}{p}}{\oper{\rho}_p(\tau)}{\m{}{p+1}}
     ,
\end{align}
for $\mathcal{V}_{l} \equiv \mathcal{V}_{l,\text{odd}} + \mathcal{V}_{l,\text{even}}$. Then, by defining $\mathcal{V}_{\text{all}} \equiv \frac{1}{L} \sum_{l=1}^L \mathcal{V}_l$, Eq. \eqref{eq:dKdg} becomes
\begin{align} 
    \label{eq:dKdg_final}
  \dv{\expval{\K{}}}{g} 
    =& -\beta \expval{\K{}} \frac{1}{\Z} \sum_{\{\m{}{l}\}_{2L}} \mathcal{V}_{\text{all}} \prod_{p=1}^{2L} \mel*{\m{}{p}}{\oper{\rho}_p(\tau)}{\m{}{p+1}}
    +\beta \frac{1}{\Z} \sum_{\{\m{}{l}\}_{2L}} \mathcal{K} \cdot \mathcal{V}_{\text{all}} \prod_{p=1}^{2L} \mel*{\m{}{p}}{\oper{\rho}_p(\tau)}{\m{}{p+1}}
    \nn
    =& \beta \left[\expval{\mathcal{K} \cdot \mathcal{V}_{\text{all}}} - \expval{\K{}}\cdot\expval{\V{}}_{\text{all}} \right]
    .
\end{align}

\section{Computational details for the MCMC algorithm}

\begin{figure*}[hb]
\includegraphics[width = 0.9\textwidth]{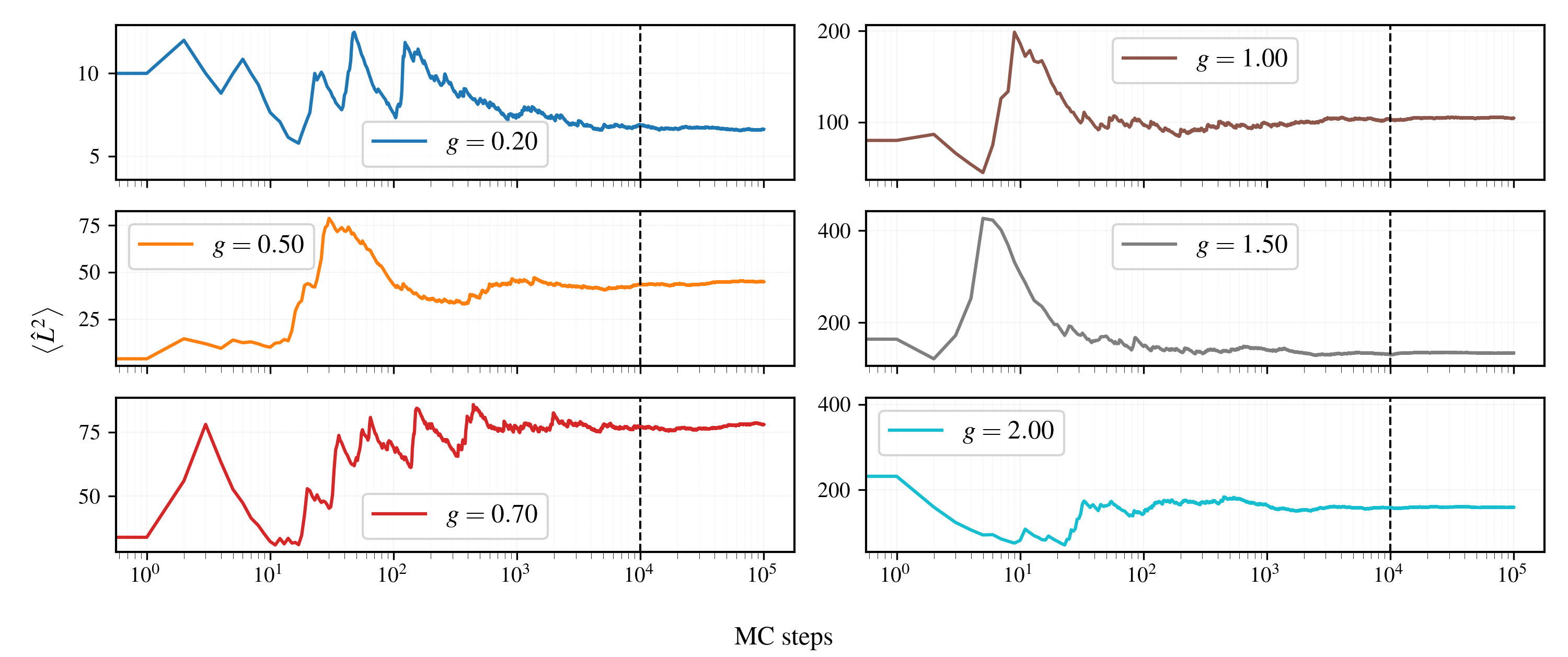}
\vspace{-0.3cm}
\caption{\label{fig:L2_vs_MC} Average total angular momentum squared $\expval{\oper{L}^2}$ as a function of simulation steps for different dipole-dipole interaction strengths $g$, for $N=150$ and $\mmax=5$. The black dotted line sets the equilibration time $10^4$ chosen in the data process procedure.}
\end{figure*}

A system of $N=150$ planar rotors with $\mmax=5$ was simulated for different values of the dipole-dipole interaction strength $g$, above and below the estimated critical interaction strength of $g_c \sim 0.5$. For all  the simulations, a total of $10^5$ MC steps were used. In order to insure faster equilibration, the initial simulation started at a high $g$ with a random initial state, and the next simulations for lower $g$'s had their grid state initialized with the final configuration of the previous simulation. During the data processing procedure the average total angular momentum squared $\expval{\oper{L}^2}$ as a function of simulation steps was calculated for different dipole-dipole interaction strengths $g$ (see Fig. \ref{fig:L2_vs_MC}) for the purpose of setting the equilibration time.


\end{document}